\documentclass[aps, pre, onecolumn, nofootinbib, notitlepage, groupedaddress, amsfonts, amssymb, amsmath, longbibliography]{revtex4-1}
\usepackage{CJK}
\usepackage{graphicx}
\usepackage{hyperref}
\usepackage{xcolor}
\hypersetup{
    colorlinks,
    linkcolor={red!50!black},
    citecolor={blue!50!black},
    urlcolor={blue!80!black}
}
\usepackage{bm}
\usepackage{natbib}
\usepackage{longtable}
\usepackage{enumitem}
\setlist[description]{leftmargin=\parindent,labelindent=\parindent}

\usepackage{siunitx}

\newcommand{\approptoinn}[2]{\mathrel{\vcenter{
	\offinterlineskip\halign{\hfil$##$\cr
	#1\propto\cr\noalign{\kern2pt}#1\sim\cr\noalign{\kern-2pt}}}}}

\begin{document}
\begin{CJK*}{GB}{}
	
\author{Zhe Feng}
\affiliation{Department of Mechanical Engineering, National University of Singapore, 9 Engineering Drive 1, 117575 Singapore}
\author{Dongdong Wan}
\affiliation{Department of Mechanical Engineering, National University of Singapore, 9 Engineering Drive 1, 117575 Singapore}
\author{Bo-Fu Wang}
\affiliation{Shanghai Key Laboratory of Mechanics in Energy Engineering, Shanghai Institute of Applied Mathematics and Mechanics, School of Mechanics and Engineering Science, Shanghai University, Shanghai 200072, China}
\author{Mengqi Zhang}
\affiliation{Department of Mechanical Engineering, National University of Singapore, 9 Engineering Drive 1, 117575 Singapore}

\title{Nonlinear spatiotemporal instabilities in two-dimensional electroconvective flows}

\begin{abstract}
This work studies the effects of a through-flow on two-dimensional electrohydrodynamic (EHD) flows of a dielectric liquid confined between two plane plates, as a model problem to further our understanding of the fluid mechanics in the presence of an electric field. The liquid is subjected to a strong unipolar charge injection from the bottom plate and a pressure gradient along the streamwise direction (forming a Poiseuille flow). Highly-accurate numerical simulations and weakly nonlinear stability analyses based on multiple-scale expansion and amplitude expansion methods are used to unravel the nonlinear spatiotemporal instability mechanisms in this combined flow. We found that the through-flow makes the hysteresis loop in the EHD flow narrower. In the numerical simulation of an impulse response, the leading and trailing edges of the wavepacket within the nonlinear regime are consistent with the linear ones, a result which we also verified against that in natural convection. In addition, as the bifurcation in EHD-Poiseuille flows is of a subcritical nature, nonlinear finite-amplitude solutions exist in the subcritical regime, and our calculation indicates that they are convectively unstable (at least for the parameters investigated). The validity of the Ginzburg-Landau equation (GLE), derived from the weakly nonlinear expansion of Navier-Stokes equations and the Maxwell's equations in the quasi-electrostatic limit, serving as a physical reduced-order model for probing the spatiotemporal dynamics in this flow, has also been investigated. We found that the coefficients in the GLE calculated using amplitude expansion method can predict the absolute growth rates even when the parameters are away from the linear critical conditions, compared favourably with the local dispersion relation, whereas the validity range of the GLE derived from the multiple-scale expansion method is confined to the vicinity of the linear critical conditions.
\end{abstract}
\maketitle
\end{CJK*}

\section{Introduction} \label{sec:intro}
In the current work, the electroconvection (EC) in a dielectric liquid with a low electrical conductivity is studied. As the induced electric current is weak in the dielectric liquid, the magnetic field can be neglected. Ions can be generated via different methods (to be discussed below) and injected into the liquid. Under a DC (i.e. direct current) electric field, the injected ions can give rise to Coulomb force, which can influence and control the motion of liquids in various industrial applications. The electrohydrodynamic (EHD) conduction pumping technique can be used to electrically drive the dielectric liquid flow in a single-phase loop \citep{feng2004understanding}. The charged ions are generated in a dissociation process of the neutral electrolytic species in the vicinity of electrodes. Aside from controlling the mass transfer in dielectric liquids, the combined effects of an electrical field and buoyancy can enhance the efficiency of heat transfer \citep{mccluskey1991heat}. For two miscible dielectric liquids under an electric field, the Coulomb force can destabilise the interface by inducing fluid circulations around the interface \citep{Ward2019}. The experimental results of Ref. \citep{jalaal2013electrohydrodynamic} indicate that such EHD mixers can significantly increase the efficiency of mixing in liquid-liquid systems. 

When the electroconvective flow is subjected to a through-flow, it can also find numerous applications in various industrial processes. For example, Atten \& Honda \cite{atten1982electroviscous} have investigated the EHD-Poiseuille flow to understand the electroviscous (EV) phenomenon (referring to the increase of apparent viscosity in a bulk flow as a result of the electrostatic effect). The EV effect can be utilised to damp undesired vibration of devices in some particular circumstances and such vibration dampers are useful in the case where conventional mechanical dampers are inapplicable.  Another relevant application is the electrostatic precipitator (ESP), where a through-flow with dusts and particles interacts with the electric field confined between two plates with high-voltage wires in-between. ESPs serve as an effective and reliable particulate control device for cleaning industrial and polluted gases and can reduce the unwanted particles of various sizes \cite{soldati1993direct,parker1997electrostatic,mizuno2000electrostatic,Soldati2000,soldati2003cost}. Some ESP can efficiently remove particles larger than $\SI{2.5}{\micro\metre}$, but the collection efficiency decreases for smaller particles ($<\SI{2.5}{\micro\metre}$) \cite{jaworek2018two}. Currently, we cannot find an effective solution to this problem probably due to a lack of a systematic understanding of how the electrified flow interacts with the through-flow.

In this work, we will study the problem of how a through-flow interacts with the flow induced by an electric field from the perspective of flow instability. Such knowledge can potentially in the end help us to, for example, enhance the efficiency of ESP, design more efficient vibration damper etc., as mentioned above. In order to analyse the physical problem with only the most essential elements, in this work we simplify the problem by considering the spatiotemporal instability of a Poiseuille flow subjected to an electric field in a dielectric liquid confined between two infinite plane walls. The ions are generated via the physicochemical process near the ion-exchange membrane covering the walls. In subsequent work where a more similar flow configuration to ESP is considered, the results may be applicable to the flows in ESP more straightforwardly; we study the aforementioned academic idealised flow in this work. In following sections, the linear and nonlinear instabilities of this EHD flow with and without the through-flow, as well as the linearly and nonlinearly convective and absolute instabilities of the EHD flow and other flows, are reviewed.

\subsection{Linear and weakly nonlinear instabilities of EHD flows}
The EHD flow studied here is governed by the Navier-Stokes equations for the flow field and the reduced Maxwell's equation in the quasi-electrostatic limit for the electric field \cite{castellanos1998electrohydrodynamics}. In the case of space-charge-limit (SCL) injection (referring to an infinitely strong level of charge injection), the linear stability criterion only depends on the electric Rayleigh number $T$ \cite{Schneider1970,atten1972stabilite}, which is a dimensionless number quantifying the strength of external voltage added on the top and bottom plates (see the problem formulation section \ref{sec:formulation} to follow). For the EHD flow without a through-flow, Atten \& Moreau \cite{atten1972stabilite} determined the linear stability criterion $T_c \approx 160.75$ (which has also been subsequently confirmed in similar linear stability analyses \citep{wu2015two,zhang2015modal,wang2016numerical}). However, in the EHD experiments using ion-exchange membranes to generate ions \citep{lacroix1975electro,atten1978electrohydrodynamic}, the onset of electroconvection took place at about $T_c \approx 100 \pm 10$. Continuous efforts have been devoted to solving this now long-standing problem in the EC flows from several perspectives, including the effect of neglected charge diffusion, the non-modal energy growth and the validity of homogeneous charge injection assumption. P\'erez \& Castellanos \cite{perez1989role} considered the non-negligible effect of the charge diffusion on the linear stability criterion and found that the value of $T_c$ decreases with a stronger charge diffusion effect. Zhang et al. \cite{zhang2015modal} analysed the modal and non-modal behaviours of this linearised EHD flow, but found that the transient growth therein is small. Recently, Feng et al. \cite{feng2021deterministic} have considered the possible deleterious effect of the inhomogeneity on ion exchange membranes, which is inevitable in experiments but is conventionally neglected in the stability analyses of EHD flows, by introducing some stochasticity in the boundary conditions of their numerical simulations. Their results showed that with stronger stochasticity, the electroconvection can indeed take place at a lower electric Rayleigh number (rendering the numerical prediction of $T_c$ closer to the experimental value).

In the case of EHD-Poiseuille flows, the combined effects of the through-flow and the secondary flow induced by electric field have also been investigated. This flow was experimentally studied by Atten \& Honda \cite{atten1982electroviscous}. At the onset of the EHD secondary motions, the apparent viscosity was found to increase which can be indicated by the increased pressure gradient when keeping a constant flow rate. The linear stability analysis of EHD-Poiseuille flows was performed by Castellanos \& Agrait \cite{Castellanos1992}. At low Reynolds numbers ($Re$), the most unstable modes are always transverse rolls, and the transverse perturbations are stabilised. But they are destabilised at high $Re$. As the decrease of ratio between hydrodynamic and ionic mobilities, transverse rolls are also destabilised in the case of low Reynolds numbers \cite{lara1997destabilization}. The modal and non-modal stabilities of EHD-Poiseuille flow were studied by Zhang et al. \cite{zhang2015modal}. They found that the linear instability criterion decreases compared with the no-through-flow case, and the transient growth of energy increases with the intensification of electric fields, which indicates that Poiseuille flows can be destabilised by the electric field in a short time.

Beyond the linear instability $T_c$ in EHD flows, an infinitesimal disturbance will continue to grow and the nonlinearity will start to manifest itself when the disturbance amplitude is large enough. At this point, the nonlinear effects should be analysed. A weakly nonlinear stability analysis is an elegant mathematical tool that can unravel the effect of weak nonlinearity in an analytical manner. This analysis approach was first conceptually proposed by Landau \cite{Landau1944}, who, based on an argument of physical symmetry, proposed the amplitude equation governing the evolution of disturbance around the primary bifurcation at the linear criticality. When the diffusion effect is considered in the derivation of the amplitude equation, one can obtain the Ginzburg-Landau equation (GLE). In the context of hydrodynamic stability, the GLE was first derived rigorously by Stuart using a multiple-scale expansion method \cite{Stuart1960Nonlinear} and by Watson using an amplitude expansion method \cite{Watson1960Nonlinear}. These two methods have been proved to be equivalent to each other at the critical point later \cite{Fujimura1989Equivalence}. The weakly nonlinear analysis has been applied successfully in canonical shear flows and Rayleigh-B\'enard convection \cite{Cross1993Pattern}, etc. In EHD flows, similar efforts have also been made to understand their nonlinear effects using the weakly nonlinear stability theory, as reviewed in the following.

In a study of two-dimensional EHD flow using a hydraulic model, Felici \cite{Felici1971DC} found two equilibrium solutions (one is of zero amplitude and the other finite amplitude) at $T<T_c$, directly proving the existence of a subcritical bifurcation in the EHD flow. The results of subcritical bifurcation have been theoretically corroborated by Worraker \& Richardson \cite{Worraker1981Nonlinear} in their weakly nonlinear analysis of thermally stabilized EHD flows (the flow becomes isothermal EHD flow when the thermal difference between two plates is zero). A more systematic weakly nonlinear analysis for the EHD flow has been conducted recently by Zhang \cite{zhang2016weakly}, who investigated the primary bifurcation of EHD with and without through-flows using the multiple-scale expansion method. His results show the universality of the subcritical bifurcation in a large parameter space. This analysis is confined in the vicinity of the linear critical conditions. The weakly nonlinear behaviour of the EHD flow away from the linear criticality also deserves to be investigated, which can be performed using the amplitude expansion method (in which the linear growth rate is not necessarily small). It is noted that the amplitude expansion method has recently been improved by Suslov and his coworker \cite{Pham2018Definition} (see also Ref. \cite{Cudby2021Weakly}) to be more rigorous and versatile and this method will be utilised in the current work to probe the EHD flow when it is away from the linear criticality.

\subsection{Linear and nonlinear convective/absolute instabilities}
When an infinitesimal impulse is added to a through-flow, the resultant disturbance (taking the form of a wavepacket, that is, disturbance with a broad spectrum structure) will grow in both temporal and spatial directions. In which direction it grows faster is an important question in the spatiotemporal stability analysis. In the linear phase, the concepts of linear absolute (LA) and convective (LC) instabilities were first proposed by Briggs \cite{briggs1964electron}, among others, in the study of plasma instabilities. Later, this methodology was adopted by the researchers in the field of hydrodynamic instability, most notably Huerre, Monkewitz and Chomaz \citep{huerre1985absolute,huerre1990local,Chomaz2005} among many others. If the initial small-magnitude wavepacket grows with time but is convected away from the source location (so that the flow will eventually be stable at all positions), the flow is convectively unstable, and such disturbance cannot contaminate the upstream region. If the disturbance is amplified and extends in both upstream and downstream directions, the flow is called absolutely unstable.

By deriving the dispersion relation from the linearised governing equations, Carri\`ere \& Monkewitz \cite{carriere1999convective} studied the convective and absolute instabilities in Rayleigh-B\'enard-Poiseuille (RBP) convection of a Newtonian fluid, which is constrained between two infinite parallel plates and subjected to a thermal buoyancy force and external pressure gradient. In their work, an impulse disturbance was imposed on the linearised flow (governed by the linearised Navier-Stokes equations), which mathematically corresponds to adding a force in the form of the Dirac delta function to the linearised equations. The solution to the perturbed linearised equations is then their Green's function. Using the steepest decent method \citep{huerre1985absolute}, Carri\`ere \& Monkewitz can approximate the long-time behaviour of the Green's function for each propagation ray and thus were able to determine the absolute/convective instabilities of the disturbance. Besides, because of the periodicity of the disturbance in the streamwise and spanwise directions in their problem, Fourier transform can be applied in these directions and thus each disturbance mode (with a specified combination of streamwise and spanwise wavenumbers) can be studied separately for its absolute growth rate. They found that the disturbance mode reaching zero group velocity at the convective-absolute transition always corresponds to transverse rolls, while the disturbance region occupied by the pure longitudinal rolls will remain convectively unstable for all non-zero Reynolds numbers. This algorithm was later applied to analyze the linear convective/absolute instabilities in EHD-Poiseuille flows by Li \textit{et al.} \citep{Li2019}. Their results confirmed the transition from the convective instability to absolute instability in the region occupied by the transverse rolls when the electric Rayleigh number increases. The growth rate of oblique rolls is smaller than that of transverse rolls. These conclusions regarding the linear absolute/convective instabilities on EHD appear to be similar to those of RBP. In this work, we will explore the similarities and dissimilarities of EHD-Poiseuille and RBP flows in the nonlinear regime.


In addition to constructing above dispersion relations to study the spatiotemporal instability of a flow, the large-time asymptotic response of an impulse can also be retrieved from the results of numerical simulations by a post-processing procedure \citep{delbende1998absolute,delbende1998nonlinear}. This method is proper for analysing nonlinear absolute and convective instabilities of finite-amplitude disturbance. In this case, LA and LC are replaced by their nonlinear counterparts, namely the nonlinear absolute (NA) and convective (NC) instabilities which were first systematically studied by Chomaz \cite{chomaz1992absolute}. If the system relaxes to the basic state everywhere in the laboratory frame for all initial disturbances of finite extent and finite amplitude, the instability is nonlinearly convective. While if the system is not able to relax to the basic state everywhere in the laboratory frame for some initial condition of finite extent and amplitude, the instability is said to be nonlinearly absolute. To demonstrate the physical relevance and significance of these nonlinear concepts, Chomaz studied the one-dimensional subcritical GLE in a semi-infinite domain, and the hysteresis loop was shown to be restricted to the nonlinear absolute instability range of the control parameter \cite{chomaz1992absolute}. Later, the nonlinear convective and absolute instabilities of parallel plane wakes were studied in the system of two-dimensional Navier-Stokes equations using highly accurate numerical simulations \citep{delbende1998nonlinear}. In the subcritical parameter range of G\"ortler vortices with the asymptotic suction boundary layer over concave walls, its instability is proved to be nonlinearly convective by both theoretical study \cite{park1988convective} and experiment \cite{chomaz1991nature}. The front propagation of the linear wavepacket was calculated in this work, and it was found that the nonlinearity can drive the amplitude of linear impulse to saturate within the wavepacket as the system evolved from linear to nonlinear regimes, but the trailing and leading edges of the wavepacket were not affected, namely the trailing and leading velocities coincide in linear and nonlinear regimes. From the viewpoint of wave propagation, the linear/nonlinear convective and absolute instabilities can also be discriminated from the signs of trailing and leading velocities. If they have opposite signs, the system is absolutely unstable. Otherwise, it is convectively unstable. In the end, the fully nonlinear dynamics of the parallel wakes, including their speed selection of front, absolute and convective instabilities, and global bifurcations were studied in the work of Chomaz \cite{chomaz2003fully}. 

\subsection{The position and structure of current work}
In this work, we aim to study the nonlinear convective and absolute instabilities in the EHD-Poiseuille flow following the methodology developed by Chomaz and his coworkers in Refs. \cite{delbende1998absolute,delbende1998nonlinear,chomaz2003fully}. As explained above, knowledge on this idealised flow can be potentially useful for the industrial applications, such as ESP \cite{soldati2003cost} and EV damper devices \cite{atten1982electroviscous}. This idealised flow retains the most salient flow dynamics in the EHD flow subjected to a through-flow and at the same time reduces the modelling difficulty in geometry, which is beneficial for nonlinear analyses. This manuscript is closely related to and extends our previous works Li et al. \cite{Li2019} on the LA/LC and Zhang \cite{zhang2016weakly} on the derivation of the GLE for describing the nonlinear dynamics in EHD-Poiseuille flows. Our extension focuses on analysing the nonlinear effects in the absolute instability/convective instability (or in short AI/CI) of EHD-Poiseuille flows using the numerical simulation and testing the relevance of nonlinear GLE (derived from a weakly nonlinear expansion of the Navier-Stokes equations and the Maxwell's equations in the quasi-electrostatic limit) in depicting the nonlinear AI/CI, following and supplementing the similar theoretical research conducted by Suslov \& Paolucci \cite{SUSLOV1997}.

The paper is organised as follows. In section \ref{sec:formulation}, we introduce the nonlinear governing equations with boundary conditions and their corresponding linearisation around the base flow, the framework for the weakly nonlinear analysis based on the multiple-scale expansion and amplitude expansion methods, as well as the numerical method used to retrieve the large-time asymptotics along a certain spatiotemporal ray from snapshots of numerical simulations. In section \ref{sec:results}, the numerical code is validated against existing results in the literature. Our results on the nature of first transition, the weakly nonlinear analysis, as well as the linear and nonlinear impulse responses are presented in detail in this section. Finally, in section \ref{sec:conclusion}, we summarise our conclusions and discuss possible future works.

\section{Problem Formulation} \label{sec:formulation}
\subsection{Nonlinear equations} \label{sec:nonlinear_equations}
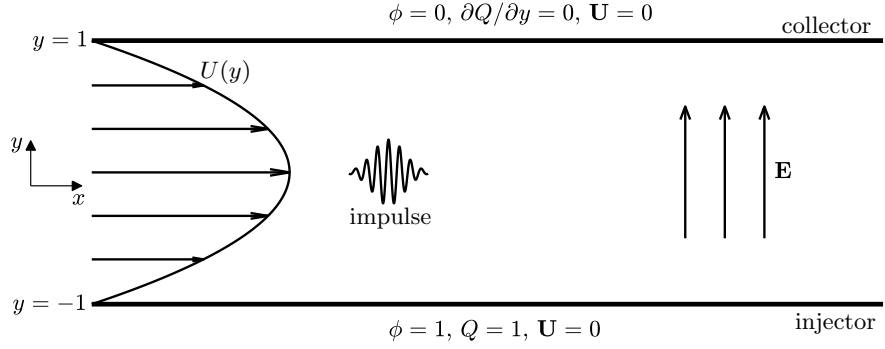
\begin{figure}
    \centering
    \includegraphics[width=0.65\textwidth]{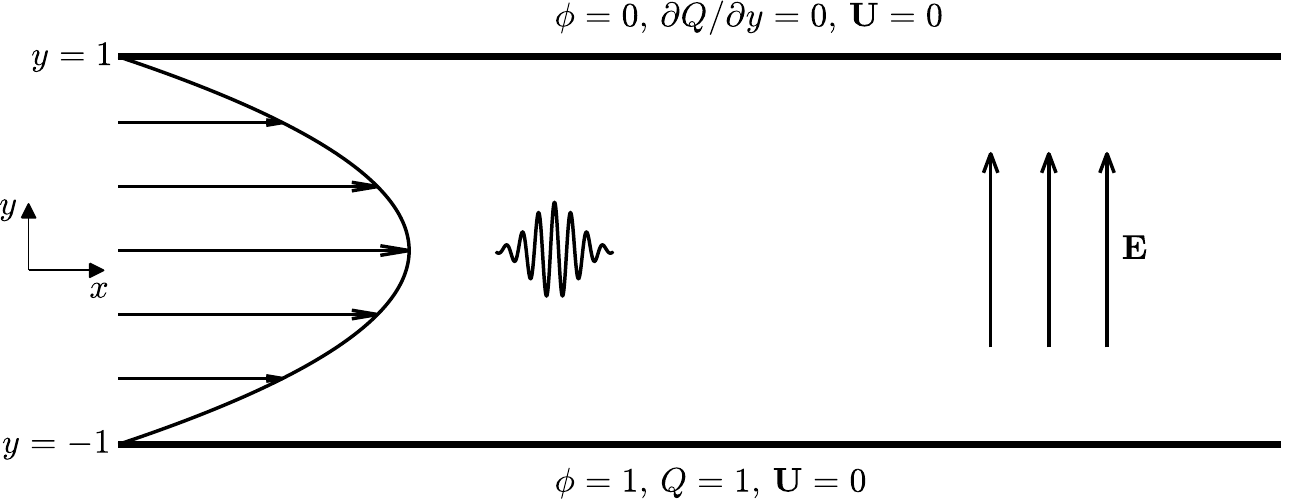}
    \put(-203.5,44.5){impulse}
    \put(-35,5){injector}
    \put(-40,117){collector}
    \put(-260,100){$U(y)$}
    \caption{Schematic of the EHD-Poiseuille flow with an initially localized impulse in the dielectric fluid which is confined between two infinite parallel plates. Charged ions are injected into the fluid from bottom plate (i.e. injector), and they are collected by top plate (i.e. collector). }
   \label{fig.schematic}
\end{figure}

The 2D EHD-Poiseuille (or EHD-P) flow that we will study in this work is confined between two parallel plates with a vertical half height $H^*$ and a horizontal length $L^*$ (the superscript asterisk $^*$ denotes dimensional variables) in a $x$-$y$ plane as shown in Fig.  \ref{fig.schematic} (with $\hat{\mathbf{x}}, \hat{\mathbf{y}}$ being unit vectors). A streamwise constant pressure gradient is imposed to drive the flow in the lateral direction. The unipolar charges are injected autonomously from the bottom plate with charge density $Q_0^*$. An external DC voltage $\Delta \phi_0^*$ is applied to the two plates. With the above dimensional variables, as well as the ionic mobility $K^*$, the permittivity $\epsilon^*$ and density $\rho_0$ of fluid, we can use the characteristic quantities $H^*$, $H^{*2}/(K^*\Delta\phi_0^*)$, $K^*\Delta\phi_0^*/H^*$, $\rho_0^*K^{*2}\Delta\phi_0^{*2}/H^{*2}$, $\Delta \phi_0^*$, $\Delta\phi_0^*/H^*$ and $Q_0^*$  to nondimensionalise the length, time, velocity, pressure, electric potential, electric field and charge density, respectively. Then the nondimensionalised governing equations of the system can be given as
\begin{subequations}\label{eq.nlinehd}
  \begin{gather}
    \nabla \cdot \mathbf{U} = 0, \\
    \frac{\partial \mathbf{U}}{\partial t} + (\mathbf{U} \cdot \nabla)\mathbf{U} = -\mathbf{\nabla} P+\frac{M^2}{T}\nabla ^2 \mathbf{U}+CM^2 Q \mathbf{E}, \\
    \frac{\partial Q}{\partial t} + \nabla \cdot [(\mathbf{E}+\mathbf{U})Q] = \frac{1}{Fe}\nabla ^2 Q,  \\
    \nabla ^2 \phi = -CQ, \\
    \mathbf{E} = -\nabla \phi, \\
    \text{with \ \ \ \ } M=\frac{(\epsilon^*/\rho_0^*)^{1/2}}{K^*}, \,\, T=\frac{\epsilon^*\Delta \phi_0^*}{K^*\mu^*}, \,\, C=\frac{Q_0^*H^{*2}}{\Delta \phi_0^* \epsilon ^*}, \,\, Fe=\frac{K^*\Delta \phi_0^*}{D_{\nu}^*},  \nonumber
  \end{gather}
\end{subequations}
where the dielectric liquid is assumed to be incompressible, Newtonian and isothermal. The magnetic field is neglected in the reduced set of Maxwell's equations due to the low electric conductivity of the dielectric liquid \citep{castellanos1998electrohydrodynamics}. Regarding the normalisation method, in the case of turbulent EHD-Poiseuille flows, previous works \cite{Leonard1983,soldati1998turbulence} used the shear velocity as the characteristic velocity. The current work intends to study the spatiotemporal instability of EHD-Poiseuille flows, thus the control parameters are not strong enough to trigger turbulence. The dimensionless parameter $M$ denotes the ratio between the hydrodynamic mobility $(\epsilon^*/\rho_0^*)^{1/2}$ and the ionic mobility $K^*$ with $M<0.1$ for gases and $M$ at least of order one for liquids \citep{Castellanos1992}. The ratio of the Coulomb force to the viscous force is described by the electric Rayleigh number $T$, which plays a similar role as the Rayleigh number $Ra$ in RBC (short for Rayleigh-B\'enard convection). The third parameter $C$ quantifies the injection strength of charges, and $C>1$ indicates strong injection, including the SCL injection, which is the scenario considered in this work. The last dimensionless parameter $Fe$ represents the reciprocal of the charge diffusivity coefficient, and it typically ranges from $10^3$ to $10^4$ \citep{perez1989role}. The boundary conditions for field variables $\mathbf{U}=(U,V)$, $Q$ and $\phi$ are
\begin{subequations}\label{eq.nlinehdBC}
  \begin{align}
    y=-1&: \; \mathbf{U}=0, \, Q=1, \, \phi=1; \\
    y=1&: \; \mathbf{U}=0, \, \partial Q/\partial y=0, \, \phi=0; \\
    x=0 \; \& \; x=L&: \; \text{periodic boundary conditions}.
  \end{align}
\end{subequations}

A streamwise constant pressure gradient is provided to sustain the base Poiseuille flow $U(y)=\mathcal{U}(1-y^2)$ (with $y\in[-1,1]$), where $\mathcal{U}$ denotes the maximum velocity magnitude of the Poiseuille flow, and it can be used to control the strength of the through-flow. As $U(y)$ is the steady solution to the momentum equation (\ref{eq.nlinehd}b) with no-slip boundary conditions, the constant pressure gradient can be calculated as
\begin{equation}\label{eq.prgrad}
  U(y) = -\frac{1}{2}\frac{dP}{dx}\frac{T}{M^2}\left(1-y^2\right),\,\ \frac{dP}{dx} = - 2\mathcal{U}\frac{M^2}{T}.
\end{equation}

To evaluate the electric transfer in EHD-Poiseuille flows, the following electric Nusseult number $Ne$ \citep{zhang2016weakly} is defined
\begin{equation} \label{nedef}
  Ne=\frac{I_e}{I_0},  \; I_e=\iint_{\mathcal{V}} \bigg[Q(V+E_y) - \frac{1}{Fe}\frac{\partial Q}{\partial y} \bigg]dxdy,
\end{equation}
where $Ne$ denotes the ratio between total current $I_e$ (with convective motions, i.e., $\mathbf{U}\neq 0$) and the conductive current $I_0$ (without convective motions, i.e., $\mathbf{U}=0$).

\subsection{Perturbative form of the governing equations}
In order to study the linear and nonlinear instabilities in EHD-Poiseuille flows, the field variables are decomposed into a summation of base states (which are steady solutions to Eqs. \ref{eq.nlinehd}) and perturbations, i.e. $\mathbf{U}=\bar{\mathbf{U}}+\mathbf{u}$, $P=\bar{P}+p$, $\mathbf{E}=\bar{\mathbf{E}}+\mathbf{e}$, $Q=\bar{Q}+q$ and $\phi=\bar{\phi}+\varphi$, which can be substituted into the nonlinear governing equations (\ref{eq.nlinehd}) to yield
\begin{subequations}\label{eq.linehd}
  \begin{gather}
    \nabla \cdot \mathbf{u} = 0,  \\
    \frac{\partial \mathbf{u}}{\partial t}+(\mathbf{u}\cdot \nabla)\bar{\mathbf{U}} + (\bar{\mathbf{U}}\cdot \nabla) \mathbf{u} = -\nabla p + \frac{M^2}{T}\nabla ^2 \mathbf{u} + CM^2(q\bar{\mathbf{E}}+\bar{Q}\mathbf{e}) + \mathbf{N}_{\mathbf{u}}, \\
    \frac{\partial q}{\partial t}+ \nabla \cdot [(\bar{\mathbf{E}}+\bar{\mathbf{U}})q+(\mathbf{e}+\mathbf{u})\bar{Q}] = \frac{1}{Fe}\nabla^2 q + N_q, \\
    \nabla^2 \varphi = -Cq,  \\
    \mathbf{e} = -\nabla \varphi,  
  \end{gather}
\end{subequations}
where the perturbation velocity $\mathbf{u}=(u,v)$ and electric field $\mathbf{e}=(e_x,e_y)$, $\mathbf{N}_{\mathbf{u}}=-(\mathbf{u}\cdot \nabla)\mathbf{u}+CM^2(q\mathbf{e})$ and $N_q=-\nabla \cdot [(\mathbf{e}+\mathbf{u})q]$ are nonlinear terms of perturbations. Discarding the nonlinear terms directly leads to the linearised governing equations for the perturbations. $\bar{\mathbf{E}}$ and $\bar{Q}$ refer to hydrostatic solutions of the electroconvective flow (see, e.g., Ref. \cite{zhang2015modal} for their profiles), and as mentioned above $\bar{\mathbf{U}}=\mathcal{U}(1-y^2)\hat{\mathbf{x}}$ is the Poiseuille flow. The homogeneous boundary conditions are applied for the perturbed fields 
\begin{subequations}\label{eq.linehdBC}
  \begin{align}
    y=-1&: \; \mathbf{u}=0, \, q=0, \, \varphi=0; \\
    y=1&: \; \mathbf{u}=0, \, \partial q/\partial y=0, \, \varphi=0;\\
    x=0 \; \& \; x=L&: \; \text{periodic boundary conditions}.
  \end{align}
\end{subequations}


\subsection{Numerical analysis of flow instability}
\subsubsection{Temporal instability}
The temporal instability analysis of a flow concerns the development of the disturbance on the flow in time. Time series of the disturbance can be obtained either experimentally or numerically, based on which, useful information, such as the most dangerous wavelength/wavenumbers and their associated growth rates, can be calculated in the post-processing analysis \citep{delbende1998nonlinear}.
In order to get these pieces of spectral information, Fast Fourier Transform (FFT) is conventionally applied; for example, in our numerical simulations to be presented, FFT is used to decompose the wavepacket with a broad spectrum structure into multiple waves within the linear regime of the flow evolution. The growth rate of each wave can be calculated using the temporal evolution of the total energy density; in the EHD-Poiseuille flow to be studied, the norm of total energy in the spectral space is defined as 
\begin{equation} \label{eq.espectrum}
  \tilde{\mathcal{E}}(\alpha,t) = \left[ \frac{1}{2} \int_{-1}^1 \left( \mathbf{u}(\alpha,y,t)^2+M^2\mathbf{e}(\alpha,y,t)^2 \right)dy \right]^{1/2}, 
\end{equation}
at streamwise wavenumber $\alpha$. The total energy density of EHD-Poiseuille flows consists of kinetic energy density $\mathcal{E}_k=\mathbf{u}^2/2$ and electric energy density $\mathcal{E}_{\varphi}=M^2\mathbf{e}^2/2$ \citep{castellanos1998electrohydrodynamics,zhang2016weakly}. When solving the nonlinear equations (\ref{eq.nlinehd}), we use the perturbed velocity and electric field in equation (\ref{eq.espectrum}) by subtracting the base states from the nonlinear fields (i.e., $\mathbf{u}=\mathbf{U}-\bar{\mathbf{U}}$ and $\mathbf{e}=\mathbf{E}-\bar{\mathbf{E}}$). In the linear regime, $\tilde{\mathcal{E}}(\alpha,t)$ presents an exponential scaling at an asymptotically large time according to the linear theory, that is,
$  \tilde{\mathcal{E}}(\alpha,t) \propto e^{\omega(\alpha)t},$ with $\omega(\alpha) \sim \frac{\partial}{\partial t} \ln \tilde{\mathcal{E}}(\alpha,t)$ as $t \rightarrow \infty$. Thus, following Ref. \cite{delbende1998nonlinear}, the leading growth rate at wavenumber $\alpha$ can be computed from the norm of total energy density by $\omega(\alpha) \approx \frac{\ln \left[ \tilde{\mathcal{E}}(\alpha,t_2)/\tilde{\mathcal{E}}(\alpha,t_1) \right]}{t_2 - t_1}$
where the wavenumber is constrained in the range $-\pi/\delta x < \alpha < \pi/\delta x$ due to the spatial discretization ($\delta x$ refers to the interval of adjacent grid points along the streamwise direction), and the smallest increment of wavenumber is $\delta \alpha = 2\pi / L$. The linear stability theory also dictates that the flow field at an asymptotically large time is proportional to the eigenvector. With this fact, we can verify our calculation by checking the invariance of the quantity $  \tilde{\mathcal{E}}_f(\alpha,y,t) = \frac{\hat{\tilde{\mathcal{E}}}(\alpha,y,t)}{\mathcal{E}(\alpha,t)}$, where the eigenvector $\hat{\tilde{\mathcal{E}}}(\alpha,y,t)$ at streamwise wavenumber $\alpha$ is normalized by its amplitude $\tilde{\mathcal{E}}(\alpha,t)$.

\subsubsection{Spatiotemporal instability}
The temporal stability analysis only concerns the development of the disturbance in time. When a wave is propagating in a medium, its development in space should also be studied. Considering both the spatial and temporal development of the disturbance helps to differentiate two types of instability mechanisms, i.e., convective instability and absolute instability. One can characterise the spatiotemporal development of an unstable mode in the physical space by the temporal growth rate $\omega$ and the spatial group velocity $v_g$; the latter refers to the propagation speed of the overall envelope of the wavepacket. As mentioned in the introduction section, theoretical methods construct a dispersion equation to probe the asymptotic spatiotemporal dynamics (see Ref. \cite{huerre1985absolute} for a general procedure or our previous work \cite{Li2019} for a more relevant reference) and post-processing methods analyse the numerical data to extract the spatiotemporal information of a flow. In this work, we adopt the latter. To unambiguously extract the amplitude and the phase of a wavepacket, the Hilbert transform is applied to the flow variables \citep{brancher1997absolute}, which reads
\begin{equation} \label{eq.hilbert}
  \tilde{g}(x,y,t) = \hat{\tilde{g}}(x,y,t) e^{i \psi(x,y,t)},
\end{equation}
where $\hat{\tilde{g}}(x,y,t)$ represents the complex-valued amplitude function of any flow variable, and $\psi(x,y,t)$ is the phase of the wavepacket. The real amplitude function $\hat{\tilde{\mathcal{E}}}(x,t)$ of energy norm can be obtained by integration along the $y$-axis
\begin{equation}
  \hat{\tilde{\mathcal{E}}}(x,t) = \left[ \frac{1}{2} \int_{-1}^1 \left( \hat{\tilde{\mathbf{u}}}(x,y,t)^2+M^2\hat{\tilde{\mathbf{e}}}(x,y,t)^2 \right) dy \right]^{1/2}.
\end{equation}
It can be shown that the real amplitude function at an asymptotically large time follows (see Ref. \citep{huerre1985absolute} for the derivation)
\begin{equation} \label{eq.stasym}
  \hat{\tilde{\mathcal{E}}}(x,t) \propto t^{-1/2} e^{\sigma(v_g)t} \ \ \ \ \text{with} \ \ \ \, v_g=(x-x_0)/t=\text{const}, \, t\rightarrow \infty,
\end{equation}
where $x_0$ is the initial streamwise position of the impulse disturbance and $\sigma(v_g)$ refers to the dominant growth rate along the ray $v_g$. Based on snapshots extracted within the linear regime, the growth rate can be computed in the following discretized form
\begin{equation} \label{eq.sigma}
  \sigma(v_g) \approx \frac{\ln \left[ \hat{\tilde{\mathcal{E}}}(v_g t_2,t_2)/\hat{\tilde{\mathcal{E}}}(v_g t_1,t_1) \right]}{t_2-t_1} + \sigma_0(t_1,t_2), \ \, \sigma_0(t_1,t_2) = \frac{\ln(t_2/t_1)}{2(t_2-t_1)},
\end{equation}
where $\sigma_0(t_1,t_2)$ is a finite-time correction for the spatiotemporal growth rate, which arises from the $t^{-1/2}$ factor in equation (\ref{eq.stasym}) \citep{delbende1998nonlinear}.

In some numerical instances to be followed, the nonlinear equations (\ref{eq.nlinehd}) are solved to probe the spatiotemporal instability of the EHD-Poiseuille flows. It has been noticed in Refs. \cite{Melville1983,delbende1998nonlinear}  and we also observed that the Hilbert transform (\ref{eq.hilbert}) cannot accurately extract the amplitude function and the phase of nonlinearly unstable wavepackets. In this case, the enstrophy is used in the nonlinear analysis, following Refs. \cite{Melville1983,delbende1998nonlinear}, 
\begin{equation}\label{eq.eta}
  \eta(x,t) = \left( \int_{-1}^1 \omega^2(x,y,t)dy \right)^{1/2},
\end{equation}
where $\omega=\partial v/\partial x - \partial u/\partial y$ denotes the vorticity in 2D flows.

In the end, since the spatiotemporal instability properties of a flow can be inferred from its response to an impulse disturbance, some physical understanding of the impulse response is instructive. Edge velocities $v_{\pm}$ refer to those at which neutral waves propagate (i.e. $\sigma(v_{\pm})=0$), see Fig. \ref{fig.validateT190}(c) to follow, and can help discriminate between linearly convective and absolute instabilities. The leading velocity is always positive $v_{+}>0$ as the Poiseuille flow moves downwards along the $x$-axis. Thus, if we have $v_{-}>0$ and the spatiotemporal ray $v_g=0$ lies outside the range of exponentially growing modes (i.e., $\sigma(v_g=0)<0$), the flow is convective instability; otherwise, if $v_{-}<0$ and the spatiotemporal ray $v_g=0$ lies in the range of exponentially growing modes (i.e., $\sigma(v_g=0)>0$), the flow is absolutely unstable whenever it is unstable.

\subsection{Weakly nonlinear stability analysis}
In this section, we will present the theoretical derivation of the GLE in the weakly nonlinear stability analysis. It can be used as a reduced-order model to study the spatiotemporal development of a flow.
In two-dimensional EHD problems, by using the streamfunction formulation $\psi$ (with $u=\frac{\partial \psi}{\partial y}$ and $v=-\frac{\partial \psi}{\partial x}$), eliminating the pressure term, and expressing the  variables related to the electric field in terms of $\varphi$, we can recast the equation system (\ref{eq.linehd}) into the following two-variable form \cite{zhang2016weakly}
\begin{subequations}\label{eq.stream_poten}
	\begin{align}
	\frac{\partial \nabla^2 \psi}{\partial t} = &-\bar{U} \frac{\nabla^2 \psi}{\partial x} + \bar{U}'' \frac{\partial \psi}{\partial x} + \frac{M^2}{T} \nabla^4 \psi - M^2 \left(\bar{\phi}' \frac{\partial \nabla^2 \varphi}{\partial x} - \bar{\phi}''' \frac{\partial \varphi}{\partial x} \right) \notag \\
	&- \left( \frac{\partial \psi}{\partial y} \frac{\partial \nabla^2 \psi}{\partial x} - \frac{\partial \psi}{\partial x} \frac{\partial \nabla^2 \psi}{\partial y} \right) - M^2 \left(  \frac{\partial \varphi}{\partial y} \frac{\partial \nabla^2 \varphi}{\partial x} - \frac{\partial \varphi}{\partial x} \frac{\partial \nabla^2 \varphi}{\partial y} \right), \\
	\frac{\partial \nabla^2 \varphi}{\partial t} = &\bar{\phi}'''\left( \frac{\partial \varphi}{\partial y} + \frac{\partial \psi}{\partial x} \right) + \left( \bar{\phi}' \frac{\partial}{\partial y}-\bar{U}\frac{\partial}{\partial x} \right)\nabla^2 \varphi + 2 \bar{\phi}'' \nabla^2 \varphi + \frac{1}{Fe} \nabla^4 \varphi \notag \\
	&+ \left( \frac{\partial \varphi}{\partial x} - \frac{\partial \psi}{\partial y}  \right) \frac{\partial \nabla^2 \varphi}{\partial x} + \left( \frac{\partial \varphi}{\partial y} + \frac{\partial \psi}{\partial x}  \right) \frac{\partial \nabla^2 \varphi}{\partial y} + \nabla^2 \varphi \nabla^2 \varphi,
	\end{align}
\end{subequations}
and the boundary conditions are periodic in $x$ direction and $\psi(\pm 1)=\psi'(\pm 1) =0$ and $\varphi(\pm 1)=\varphi''(1)=\varphi'''(-1)=0$ in $y$ direction (and prime $'$ means the derivative in $y$ direction). This equation system can be further expressed in a compact form as
\begin{equation}\label{eq.compact-form}
\boldsymbol{M} \frac{\partial \boldsymbol{\gamma}}{\partial t} = \boldsymbol{L} \boldsymbol{\gamma} + \boldsymbol{N}, \ \  \text{or} \ \ \left( \boldsymbol{M} \frac{\partial }{\partial t} - \boldsymbol{L} \right) \boldsymbol{\gamma} = \boldsymbol{N},
\end{equation}
where $\boldsymbol{\gamma}=(\psi,\varphi)^T$, $\boldsymbol{M}$ is the weight matrix, $\boldsymbol{L}$ is the linear operator, $\boldsymbol{N}$ is the nonlinear operator, whose explicit expressions can be deduced straightforwardly by matching with equation (\ref{eq.stream_poten}), see the appendix A and Ref. \cite{zhang2016weakly}.

There are various ways to perform a weakly nonlinear analysis and the two most classical methods are the multiple-scale expansion \cite{Stuart1960Nonlinear} around the linear critical condition and the amplitude expansion \cite{Watson1960Nonlinear} which can be valid even away from the linear critical condition. They have been proved to be equivalent to each other at critical conditions \citep{Fujimura1989Equivalence}. In the following section, we will briefly describe the key steps in the two perturbation methods.

\subsubsection{Multiple-scale expansion method} \label{sec:multiple-scale expansion}
Our multiple-scale expansion method follows the schemes in Stewartson \& Stuart \cite{Stewartson1971Nonlinear} and Fujimura \cite{Fujimura1989Equivalence}, where the time $t$, streamwise spatial coordinate $x$, perturbation $\boldsymbol{\gamma}$ and the governing parameter $T$ are expanded in series of a small quantity $\epsilon$ as
\begin{subequations}\label{eq.expansion-tzgammaT}
\begin{align}
\frac{\partial}{\partial t} &= \frac{\partial}{\partial t_{0}} + \epsilon \frac{\partial}{\partial t_{1}} + \epsilon^{2} \frac{\partial}{\partial t_{2}} + O(\epsilon^{3}), \ \ \ \ \frac{\partial}{\partial x} = \frac{\partial}{\partial x_{0}} + \epsilon \frac{\partial}{\partial x_{1}} + O(\epsilon^{2}), \\
\boldsymbol{\gamma} &= \epsilon \boldsymbol{\gamma}_{1} + \epsilon^2 \boldsymbol{\gamma}_{2} + \epsilon^3 \boldsymbol{\gamma}_{3} +  O(\epsilon^{4}), \ \ \ \ \ \ \ \ \ T = T_{c} + \epsilon^{2} + O(\epsilon^{4}),
\end{align}
\end{subequations}
where $\epsilon^2=T-T_c$ is a measure of the value $T$ away from the critical $T_c$. Accordingly, the operators introduced in equation (\ref{eq.compact-form}) should also be expanded
\begin{equation}\label{eq.expansion-MLN}
\boldsymbol{M} = \boldsymbol{M}_{0} + \epsilon \boldsymbol{M}_{1} + \epsilon^{2} \boldsymbol{M}_{2} + O(\epsilon^{3}),  \ \ \ \ 
\boldsymbol{L} = \boldsymbol{L}_{0} + \epsilon \boldsymbol{L}_{1} + \epsilon^{2} \boldsymbol{L}_{2} + O(\epsilon^{3}), \ \ \ \boldsymbol{N} = \epsilon^{2} \boldsymbol{N}_{2} + \epsilon^{3} \boldsymbol{N}_{3} + O(\epsilon^{4}).
\end{equation}

Substituting equations (\ref{eq.expansion-tzgammaT}) and (\ref{eq.expansion-MLN}) into equation (\ref{eq.compact-form}) and equating terms on the same orders of $\epsilon$ give
\begin{subequations}\label{eq.epsilons}
\begin{align}
&\left( \boldsymbol{M}_{0} \frac{\partial}{\partial t_{0}} - \boldsymbol{L}_{0} \right) \boldsymbol{\gamma}_{1} = 0, \\
&\left( \boldsymbol{M}_{0} \frac{\partial}{\partial t_{0}} - \boldsymbol{L}_{0} \right) \boldsymbol{\gamma}_{2} = \left( \boldsymbol{L}_{1} - \boldsymbol{M}_{1} \frac{\partial}{\partial t_{0}} - \boldsymbol{M}_{0} \frac{\partial}{\partial t_{1}} \right) \boldsymbol{\gamma}_{1} + \boldsymbol{N}_{2}, \\
&\left( \boldsymbol{M}_{0} \frac{\partial}{\partial t_{0}} - \boldsymbol{L}_{0} \right) \boldsymbol{\gamma}_{3} = \left( \boldsymbol{L}_{2} - \boldsymbol{M}_{2} \frac{\partial}{\partial t_{0}} - \boldsymbol{M}_{1} \frac{\partial}{\partial t_{1}} - \boldsymbol{M}_{0} \frac{\partial}{\partial t_{2}} \right) \boldsymbol{\gamma}_{1} + \left( \boldsymbol{L}_{1} - \boldsymbol{M}_{1} \frac{\partial}{\partial t_{0}} - \boldsymbol{M}_{0} \frac{\partial}{\partial t_{1}} \right) \boldsymbol{\gamma}_{2} + \boldsymbol{N}_{3}.
\end{align}
\end{subequations}
Once we assume the first-order solution $\boldsymbol{\gamma}_{1}$ to be normal-mode, the solutions  $\boldsymbol{\gamma}_{2}$ and $\boldsymbol{\gamma}_{3}$  can also be deduced and they are
\begin{subequations}\label{eq.gamma1}
	\begin{align}
	&\boldsymbol{\gamma}_{1} = A_1(x_{1},t_{1},t_{2})\tilde{\boldsymbol{\gamma}}_1(y)e^{i \alpha x_{0} + \mu t_{0}} + c.c., \\
	&\boldsymbol{\gamma}_{2} = A_1^*A_1\tilde{\boldsymbol{\gamma}}_{20}(y)  + \left[\frac{ \partial A_1}{ \partial x_1}\tilde{\boldsymbol{\gamma}}_{21}(y) e^{i \alpha x_{0} + \mu t_{0}}  + c.c.\right] + \left[A_1^2 \tilde{\boldsymbol{\gamma}}_{22}(y) e^{2i \alpha x_{0} + 2\mu t_{0}} + c.c.\right], \\
	&\boldsymbol{\gamma}_{3} =\left[A_1\tilde{\boldsymbol{\gamma}}_{31,1}(y) e^{i \alpha x_{0} + \mu t_{0}}  + c.c.\right]+  \left[\frac{ \partial^2 A_1}{ \partial x_1^2}\tilde{\boldsymbol{\gamma}}_{31,2}(y) e^{i \alpha x_{0} + \mu t_{0}}  + c.c.\right] + \left[A_1 |A_1|^2 \tilde{\boldsymbol{\gamma}}_{31,3}(y) e^{i \alpha x_{0} + \mu t_{0}} + c.c.\right] + \cdots,		
	\end{align}
\end{subequations}
where $A_1$ is the complex amplitude of the perturbation, $\alpha$ the streamwise wavenumber, $\mu$ the complex frequency with its real part being the linear growth rate, and $c.c.$ the complex conjugate of the preceding one term.

Evolution equations of $A_1$ can be derived by enforcing the solvability condition \cite{bender2013advanced} on equation (\ref{eq.epsilons}b and \ref{eq.epsilons}c), i.e., the projection of the secular terms on the right hand side of these two equations on the adjoint eigenvector $\boldsymbol{\gamma}_{1}^{\dagger}$ (as introduced below in equation (\ref{eq.adjoint})) is zero. Following Ref.  \cite{zhang2016weakly}, we introduce the inner product
\begin{equation}\label{eq.inner-product-definition}
\langle \boldsymbol{f},\boldsymbol{g} \rangle = \frac{1}{L_p} \int_0^{L_p} \int_{-1}^{1} \boldsymbol{f} \cdot \boldsymbol{M}_{\text{w}} \boldsymbol{g} \, dy \, dx_{0} = {\langle \tilde{\boldsymbol{f}}, \tilde{\boldsymbol{g}} \rangle}_{s} + c.c.,
\end{equation}
where $L_p=2\pi/\alpha$ is the wavelength, $\boldsymbol{M}_{\text{w}}=\begin{bmatrix}
1 & 0\\
0 & \kappa^2
\end{bmatrix}$ is the weight matrix with $\kappa=M$. $\kappa$ also appears in the adjoint problem resulting from the integration by part process below. $\langle \cdot,\cdot \rangle_s$ is the corresponding inner product defined in the spectral space. Then the adjoint problem corresponding to equation (\ref{eq.epsilons}a) can be defined through integration by part \citep{Luchini2014Adjoint} based on the introduced inner product as
\begin{equation}\label{eq.adjoint}
\left\langle \left( \boldsymbol{M}_{0} \frac{\partial}{\partial t_{0}} - \boldsymbol{L}_{0} \right) \boldsymbol{\gamma}_{1}, \boldsymbol{\gamma}_{1}^{\dagger} \right\rangle = \left\langle \boldsymbol{\gamma}_{1}, \left( \boldsymbol{M}_{0}^{\dagger} \frac{\partial}{\partial t_{0}} - \boldsymbol{L}_{0}^{\dagger} \right) \boldsymbol{\gamma}_{1}^{\dagger} \right\rangle, \ \ \ \rightarrow \ \ \ \left( \boldsymbol{M}_{0}^{\dagger} \frac{\partial}{\partial t_{0}} - \boldsymbol{L}_{0}^{\dagger} \right) \boldsymbol{\gamma}_{1}^{\dagger} = 0.
\end{equation}
Here the explicit expressions of the adjoint operators $\boldsymbol{M}_{0}^{\dagger}$ and $\boldsymbol{L}_{0}^{\dagger}$ are given in the appendix \ref{sec:append_operators}. This adjoint problem is also an eigenvalue problem and it is solved similarly to the linear eigenvalue problem in equation (\ref{eq.epsilons}a) at the same parameter settings. Then the projection (see Ref. \cite{zhang2016weakly} for the detailed illustration of the projection process) to eliminate secular terms leads to the following equation involving the group velocity $c_g$
\begin{equation}\label{eq.cg}
\frac{\partial A_1}{\partial t_{1}} + c_{g}\frac{\partial A_1}{\partial x_{1}}=0 \ \ \ \text{with} \ \ \ c_{g} = - \frac{{\langle (\tilde{\boldsymbol{L}}_{1}^{\circ} - \mu_{c}\tilde{\boldsymbol{M}}_{1}^{\circ}) \tilde{\boldsymbol{\gamma}}_1, \tilde{\boldsymbol{\gamma}}_{1}^{\dagger} \rangle}_{s}}{{\langle \tilde{\boldsymbol{M}}_{0}\tilde{\boldsymbol{\gamma}}_1, \tilde{\boldsymbol{\gamma}}_{1}^{\dagger} \rangle}_{s}},
\end{equation}	
and the Ginzburg-Landau equation (GLE)
\begin{equation}\label{eq.GL-t2 scale}
\frac{\partial A_1}{\partial t_{2}} = a_{1}A_1 + a_{2}\frac{\partial^{2} A_1}{\partial x_{1}^{2}} + a_{3}{|A_1|}^{2}A_1,
\end{equation}
where the coefficients are
\begin{equation}\label{eq.Landau-coefficients}
a_{1} = \frac{{\langle \tilde{\boldsymbol{L}}_{2MT}\tilde{\boldsymbol{\gamma}}_1, \tilde{\boldsymbol{\gamma}}_{1}^{\dagger} \rangle}_{s}}{{\langle \tilde{\boldsymbol{M}}_{0}\tilde{\boldsymbol{\gamma}}_1, \tilde{\boldsymbol{\gamma}}_{1}^{\dagger} \rangle}_{s}}, a_{2} = \frac{{\langle (\tilde{\boldsymbol{L}}_{2}^{\circ} - \mu_{c}\tilde{\boldsymbol{M}}_{2}^{\circ} + c_{g}\tilde{\boldsymbol{M}}_{1}^{\circ})\tilde{\boldsymbol{\gamma}}_{1} + (\tilde{\boldsymbol{L}}_{1}^{\circ} - \mu_{c}\tilde{\boldsymbol{M}}_{1}^{\circ} + c_{g}\tilde{\boldsymbol{M}}_{0})\tilde{\boldsymbol{\gamma}}_{21}, \tilde{\boldsymbol{\gamma}}_{1}^{\dagger} \rangle}_{s}}{{\langle \tilde{\boldsymbol{M}}_{0}\tilde{\boldsymbol{\gamma}}_1, \tilde{\boldsymbol{\gamma}}_{1}^{\dagger} \rangle}_{s}}, a_{3} = \frac{{\langle \tilde{\boldsymbol{N}}_{3}^{\circ}, \tilde{\boldsymbol{\gamma}}_{1}^{\dagger} \rangle}_{s}}{{\langle \tilde{\boldsymbol{M}}_{0}\tilde{\boldsymbol{\gamma}}_1, \tilde{\boldsymbol{\gamma}}_{1}^{\dagger} \rangle}_{s}},
\end{equation}
where the symbol tilde $\tilde{}$ marks operators and eigenvectors in the spectral space, $\mu_c$ is the frequency at the critical condition. All the explicit expressions of these operators are provided in the appendix.

\subsubsection{Amplitude expansion method} \label{sec:amplitude_expansion}
Amplitude expansion method was first used in the fluid mechanics community by Watson \cite{Watson1960Nonlinear} to study weakly nonlinear flows and has recently been refined by Pham \& Suslov \cite{Pham2018Definition}. The following version of the amplitude expansion method is due to the latter. When the parameter of interest is away from the linear critical condition, the above multiple-scale expansion method is not guaranteed to be valid as the small quantify $\epsilon$ introduced in equation (\ref{eq.expansion-tzgammaT}) may be too large for the truncated power series to be converged. In the amplitude expansion method, the magnitude of the perturbation itself is taken as the small quantity, based on which the expansion reads
\begin{equation}\label{eq.amplitude_exp}
	\boldsymbol{\gamma}= [A\tilde{\boldsymbol{\gamma}}_{11}E+ c.c.]  + |A|^2 \tilde{\boldsymbol{\gamma}}_{20} + [\frac{\partial A}{\partial z}\tilde{\boldsymbol{\gamma}}_{21}E+ c.c. ]  +   [A^2 \tilde{\boldsymbol{\gamma}}_{22}E^2 + c.c.] + [A|A|^2 \tilde{\boldsymbol{\gamma}}_{31}E + c.c.] + [\frac{\partial^2 A}{\partial z^2}\tilde{\boldsymbol{\gamma}}_{31z}E+ c.c. ] + \text{H. O. T.},
\end{equation}
where $A$ is the complex amplitude of the disturbance and $E=\exp(i \alpha x)$ (not to be confused with the symbol for the electric field $\mathbf{E}$). Conventionally, in previous works \citep{Watson1960Nonlinear,Fujimura1989Equivalence,Pham2018Definition,Cudby2021Weakly}, $A$ is assumed to be only a function of time. As suggested by one of the reviewers, we also consider its spatial dependence for the analysis of spatiotemporal instability in the amplitude expansion scheme. The only requirement for the above expansion to be valid is that the disturbance amplitude is sufficiently small (even though the governing parameters may be at a finite distance away from the linear critical conditions). To facilitate the description of the solution process, we define the following operators
\begin{equation}\label{eq.operators}
\tilde{\boldsymbol{M}}_{q\alpha}\tilde{\boldsymbol{\gamma}}_{pq} E^q \equiv \boldsymbol{M}(\tilde{\boldsymbol{\gamma}}_{pq}E^q), \tilde{\boldsymbol{L}}_{q\alpha}\tilde{\boldsymbol{\gamma}}_{pq} E^q \equiv\boldsymbol{L}(\tilde{\boldsymbol{\gamma}}_{pq}E^q), \mathcal{L}_{q \alpha, \sigma} \equiv\sigma \tilde{\boldsymbol{M}}_{q\alpha}-\tilde{\boldsymbol{L}}_{q\alpha},
\end{equation}
where subscripts $pq$ denote a mode with temporal amplitude of order $p$ ($A^p$) and wavenumber $q \alpha$ (note that $p,q$ are integers).
Then the following Ginzburg-Landau equation describing the temporal evolution of the amplitude arises 
\begin{equation}\label{eq.Stuart-Landau}
	\frac{\partial A}{\partial t}=-c_{gA} \frac{\partial A}{\partial x} + \sigma A + a_{2A} \frac{\partial^2 A}{\partial x^2} + K A |A|^2 + \cdots.
\end{equation}

Substituting equations (\ref{eq.Stuart-Landau}) and (\ref{eq.amplitude_exp}) into equation (\ref{eq.compact-form}), and collecting the like terms having the same order of amplitude and wavenumber, one can obtain 
\begin{subequations}\label{eq.amplitude_equations}
	\begin{align}
		\mathcal{L}_{\alpha, \sigma} \tilde{\boldsymbol{\gamma}}_{11} &= 0, \\ \mathcal{L}_{0, 2\sigma_r} \tilde{\boldsymbol{\gamma}}_{20} &=\tilde{\boldsymbol{N}}_{20}, \\
		\mathcal{L}_{\alpha, \sigma} \tilde{\boldsymbol{\gamma}}_{21} &=  c_{gA}\tilde{\boldsymbol{M}}_{1\alpha} \tilde{\boldsymbol{\gamma}}_{11}+\tilde{\boldsymbol{n}}_{21}, \\
		\mathcal{L}_{2\alpha, 2\sigma} \tilde{\boldsymbol{\gamma}}_{22} &=\tilde{\boldsymbol{N}}_{22}, \\
		\mathcal{L}_{\alpha, \sigma} \tilde{\boldsymbol{\gamma}}_{31z}  &=- a_{2A}\tilde{\boldsymbol{M}}_{1\alpha} \tilde{\boldsymbol{\gamma}}_{11}+\tilde{\boldsymbol{n}}_{31z}, \\
		\mathcal{L}_{\alpha, \sigma+2\sigma_r} \tilde{\boldsymbol{\gamma}}_{31}   &=-K\tilde{\boldsymbol{M}}_{1\alpha} \tilde{\boldsymbol{\gamma}}_{11}+\tilde{\boldsymbol{N}}_{31},
	\end{align}
\end{subequations}
where the inhomogeneous terms $\tilde{\boldsymbol{n}}_{21}=(\tilde{\boldsymbol{L}}_{1\alpha}^{\circ} - \sigma\tilde{\boldsymbol{M}}_{1\alpha}^{\circ})\tilde{\boldsymbol{\gamma}}_{11}$ and $\tilde{\boldsymbol{n}}_{31z}=(\tilde{\boldsymbol{L}}_{2\alpha}^{\circ} - \sigma\tilde{\boldsymbol{M}}_{2\alpha}^{\circ} + c_{gA}\tilde{\boldsymbol{M}}_{1\alpha}^{\circ})\tilde{\boldsymbol{\gamma}}_{11} + (\tilde{\boldsymbol{L}}_{1\alpha}^{\circ} - \sigma\tilde{\boldsymbol{M}}_{1\alpha}^{\circ} + c_{gA}\tilde{\boldsymbol{M}}_{0\alpha})\tilde{\boldsymbol{\gamma}}_{21}$. Explicit expressions of all the operators appearing here are given in the appendix. Among these six equations, $\mathcal{L}_{\alpha, \sigma} \tilde{\boldsymbol{\gamma}}_{11} = 0$ is exactly the linear eigenvalue problem corresponding to equation (\ref{eq.epsilons}a).
For $T>T_c$ near the linear critical conditions, the linear growth rate $\sigma_r>0$, and thus, the operators $\mathcal{L}_{0, 2\sigma_r}$ and $\mathcal{L}_{2\alpha, 2\sigma}$ are non-singular (even when $T \to T_c$ where $\sigma_r \to 0$), making the two corresponding equations readily solvable \cite{Pham2018Definition,Cudby2021Weakly}. It should be noted that when $T<T_c$ (with $\sigma_r<0$), there will be resonances among decaying modes and special treatments are required \citep{Suslov2008Two,Pham2018Definition}. This is not a problem in the present analysis where we consider non-negative $\sigma_r$.
Near the linear critical condition (where the linear growth rate $\sigma_r \to 0$), the operator $\mathcal{L}_{\alpha, \sigma+2\sigma_r} \to \mathcal{L}_{\alpha, \sigma}$, which is singular by construction. In this case, for the third-order solution, it should be noted that if $\tilde{\boldsymbol{\gamma}}_{31}$ is a solution, following the linear problem $\mathcal{L}_{\alpha, \sigma} \tilde{\boldsymbol{\gamma}}_{11} = 0$, we know that $\tilde{\boldsymbol{\gamma}}_{31} + C \tilde{\boldsymbol{\gamma}}_{11}$ will also be a solution with $C$ being an arbitrary complex number. In this case, a solvability condition resorting to the adjoint problem (as described in the multiple-scale expansion) along with a normalization condition \citep{Fujimura1989Equivalence}, is conventionally enforced to eliminate the ambiguity. However, away from the linear critical condition, an orthogonality condition can be utilised to make $\tilde{\boldsymbol{\gamma}}_{31}$ and $K$ uniquely determined \cite{Pham2018Definition}, that is
$\langle \tilde{\boldsymbol{\gamma}}_{11}, \tilde{\boldsymbol{\gamma}}_{31} \rangle_{\mathcal{M}} \equiv \tilde{\boldsymbol{\gamma}}_{11}^H \mathcal{M}\tilde{\boldsymbol{\gamma}}_{31}=0$,
where the superscript $H$ denotes conjugate (Hermitian) transpose and $\mathcal{M}$ is a positive definite Hermitian matrix which in the present work is the identity matrix $I$ (see Ref. \cite{Pham2018Definition} for a general discussion on the choice of the matrix $\mathcal{M}$). Such a simple choice has the advantage of using all the information in the fundamental modes $\tilde{\boldsymbol{\gamma}}_{11}$ and $\tilde{\boldsymbol{\gamma}}_{31}$. 

With the above orthogonality condition defined, following the solution procedure proposed in \cite{Crouch1993Note} and formally proved in the recent study \cite{Pham2018Definition}, the extended equation system can be arrived
\begin{equation} \label{eq.gamma31_K}
\begin{bmatrix}
	\mathcal{L}_{\alpha, \sigma+2\sigma_r} & \tilde{\boldsymbol{M}}_{1\alpha} \tilde{\boldsymbol{\gamma}}_{11}\\
	\tilde{\boldsymbol{\gamma}}_{11}^H \mathcal{M} & 0
\end{bmatrix} \begin{bmatrix}
	\tilde{\boldsymbol{\gamma}}_{31} \\ K
\end{bmatrix} = \begin{bmatrix}
	\tilde{\boldsymbol{N}}_{31} \\ 0
\end{bmatrix},
\end{equation}
which combines the solution of $\tilde{\boldsymbol{\gamma}}_{31}$ and $K$ into a single problem and can be reliably solved away from the linear critical condition. Similarly, to determine the coefficients $c_{gA}$ and $a_{2A}$ in equation (\ref{eq.amplitude_equations}c,e), we also apply the orthogonality conditions $\langle \tilde{\boldsymbol{\gamma}}_{11}, \tilde{\boldsymbol{\gamma}}_{21} \rangle_{\mathcal{M}} =0$ and $\langle \tilde{\boldsymbol{\gamma}}_{11}, \tilde{\boldsymbol{\gamma}}_{31z} \rangle_{\mathcal{M}} =0$. Thus, two more extended equation systems can be formulated
\begin{equation}\label{eq.cgA_a2A}
	\begin{bmatrix}
		\mathcal{L}_{\alpha, \sigma} & -\tilde{\boldsymbol{M}}_{1\alpha} \tilde{\boldsymbol{\gamma}}_{11}\\
		\tilde{\boldsymbol{\gamma}}_{11}^H \mathcal{M} & 0
	\end{bmatrix} \begin{bmatrix}
		\tilde{\boldsymbol{\gamma}}_{21} \\ c_{gA}
	\end{bmatrix} = \begin{bmatrix}
		\tilde{\boldsymbol{n}}_{21} \\ 0
	\end{bmatrix},
	\begin{bmatrix}
		\mathcal{L}_{\alpha, \sigma} & \tilde{\boldsymbol{M}}_{1\alpha} \tilde{\boldsymbol{\gamma}}_{11}\\
		\tilde{\boldsymbol{\gamma}}_{11}^H \mathcal{M} & 0
	\end{bmatrix} \begin{bmatrix}
		\tilde{\boldsymbol{\gamma}}_{31z} \\ a_{2A}
	\end{bmatrix} = \begin{bmatrix}
		\tilde{\boldsymbol{n}}_{31z} \\ 0
	\end{bmatrix},
\end{equation}
As reported in Ref. \cite{Cudby2021Weakly}, the linear operators here may be poorly conditioned near the critical point even if it is nonsingular but they found that it still can be inverted to give converged results. This is also the case in the present work.

\subsubsection{Relation of the two expansion methods}
The GLE (\ref{eq.GL-t2 scale}) derived in the framework of multiple-scale expansion is in the time scale $t_2$ and space scale $x_1$ (along with equation (\ref{eq.cg}) in the time scale $t_1$ and space scale $x_1$), while the GLE (\ref{eq.Stuart-Landau}) obtained using the amplitude expansion is in the time scale $t$ and space scale $x$. Therefore, to make their coefficients comparable, we combine the GLE (\ref{eq.GL-t2 scale}) with equation (\ref{eq.cg}) and transform them into
\begin{equation}\label{eq.GLE_multiple-scale_tz}
	\frac{\partial A}{\partial t} = -c_g \frac{\partial A}{\partial x}+ a_{1}\epsilon^2A + a_{2}\frac{\partial^{2} A}{\partial x^{2}} + a_{3}{|A|}^{2}A,
\end{equation}
where the perturbation amplitude is also transformed from $A_1$ to $A$ using the relationship $A=\epsilon A_1$. To make clear the physical meaning of the perturbation amplitude $A$ for the Landau coefficient $a_3$ to be uniquely determined, the linear eigenvectors ($\tilde{\boldsymbol{\gamma}}_{1}$ in the multiple-scale expansion and $\tilde{\boldsymbol{\gamma}}_{11}$ in the amplitude expansion) in the two expansion methods are normalized so that $\mathcal{A}$ denotes the global perturbation amplitude of total energy departing from the base state \citep{thompson2001kinematics}, which is defined by
\begin{equation} \label{eq.gamplitude}
  |\mathcal{A}|^2 = \frac{1}{U_{\infty}^2 \mathcal{V}} \iint_{\mathcal{V}} \frac{1}{2} \left[ \mathbf{u}^2 + M^2 \mathbf{e}^2 \right] dxdy, \ \ \ \text{with} \ \mathbf{u}=\mathbf{U}-\bar{\mathbf{U}}, \ \mathbf{e}=\mathbf{E}-\bar{\mathbf{E}},
\end{equation}
where the velocity $U_{\infty}$ and system volume $\mathcal{V}$ are applied for the purpose of nondimensionalization. In the current work, the maximum velocity magnitude of the Poiseuille flow is used for defining $U_{\infty}$.

Then the group velocity $c_g$ should be compared with $c_{gA}$. $a_1 \epsilon^2$ (or the complex frequency $\mu$ in equation (\ref{eq.gamma1}a) away from the linear critical condition) in the multiple-scale expansion should be compared with $\sigma$ in the amplitude expansion, both denoting the growth rate of perturbed amplitude $\mathcal{A}$. The coefficients $a_2$ and $a_{2A}$ share the same meaning. $a_3$ should be compared with $K$, both referring to the Landau coefficient and are calculated within the weakly nonlinear regime (i.e. not too far from the linear instability criterion $T_c$). If they are positive, then the flow transition is subcritical; otherwise, the flow transition is supercritical. We will compare the results of these two methods in the result section.

In the end, we would like to mention in passing that Soldati \& Banerjee in Ref. \cite{soldati1998turbulence} have used a triple-decomposition method \cite{Reynolds1972} to study the turbulent flows in EHD, with the flow being decomposed into a sum of a time-average component, a coherent component and the remaining turbulent component. The triple-decomposition method can be applied to fully turbulent flows (with an emphasis on revealing the flow dynamics of the coherent structures), whereas, the expansion methods in this work are traditionally applied to transitional flows transitioning from the laminar flows. 

\subsubsection{Spatiotemporal instability analysis based on GLE}\label{sec:AICI_GLE}
The linear convective and absolute instabilities of the EHD-Poiseuille flow can also be studied using the GLE. Due to its simplicity, this equation can be considered as a reduced-order model to predict the absolute instability criterion. Linearising the GLE (\ref{eq.GLE_multiple-scale_tz}) and the GLE (\ref{eq.Stuart-Landau}) results in respectively
\begin{subequations}\label{eq.linear_GLE}
	\begin{align}
		\frac{\partial A}{\partial t} &=- c_g \frac{\partial A}{\partial x} + a_{1} \epsilon^2 A + a_{2}\frac{\partial^{2} A}{\partial x^{2}} \,\,\,\, \text{(based on the multiple-scale expansion)},\\ \frac{\partial A}{\partial t}&=-c_{gA} \frac{\partial A}{\partial x} + \sigma A + a_{2A} \frac{\partial^2 A}{\partial x^2} \,\,\,\, \text{(based on the amplitude expansion)},
	\end{align}
\end{subequations}
where the coefficients $c_g$, $a_1$ and $a_2$ should be evaluated near the linear critical condition $T_c$, whereas the coefficients $c_{gA}$, $\sigma$ and $a_{2A}$ can be evaluated at parameters away (but still not too far) from the linear critical condition $T_c$. However, the critical electric Rayleigh number $T_{ca}$ beyond which the linear instability changes from convective to absolute may be not close to $T_c$, and thus, the predicted value of these coefficients (and the resulting $T_{ca}$) from the above linearised GLE may be inaccurate (see section \ref{sec:linear_AICI} for the discussion of this point and the comparison in table \ref{tab:comparison_cga1a2}). Instead, we use the following form of the linearised GLE and the corresponding dispersion relation $D(\omega,\beta)$ can be obtained by assuming the normal mode solution $\tilde{A} e^{i \beta x - i \omega t}$ \citep{Gao2013Transition,Suslov2004Stability}:
\begin{equation}\label{eq.linear_GLE1}
\frac{\partial A}{\partial t} =- c_{gT} \frac{\partial A}{\partial x} + a_{1T} A + a_{2T}\frac{\partial^{2} A}{\partial x^{2}}, \,\,\,\, D(\omega,\beta)\equiv -\omega + c_{gT} \beta + i a_{1T} - i a_{2T} \beta^2 = 0.
\end{equation}
Here, the coefficients $c_{gT}$, $a_{1T}$ and $a_{2T}$ are directly evaluated from the dispersion relation \citep{Stewartson1971Nonlinear,Suslov2004Stability} of the linear equation (\ref{eq.epsilons}a) at each $T$ of interest, i.e.,
\begin{equation}\label{eq.coeff_dispersion}
c_{gT}=i \frac{\partial \mu}{\partial \alpha}, \, \, \, a_{1T}=\mu, \, \, \, a_{2T}=-\frac{1}{2} \frac{\partial^2 \mu}{\partial \alpha^2}.
\end{equation} 
$\beta$ is the modulation wavenumber of the fundamental wave with wavenumber $\alpha_m$ at which the maximum linear growth rate is attained. The dispersion relation $D(\omega,\beta)$ has an equilibrium at $\beta_0$ given by the condition $\partial \omega / \partial \beta |_{\beta_0}=0$, which results in $\beta_0 = c_{gT}/(2i a_{2T})$. Then the absolute complex frequency $\omega_0$ and absolute growth rate $\omega_{0,i}$ (imaginary part of $\omega_0$) can be expressed as
\begin{equation} \label{eq.abs_growthrate}
\omega_0 \equiv \omega(\beta_0) = \frac{c_{gT}^2}{2i a_{2T}} + i a_{1T} + \frac{i c_{gT}^2}{4 a_{2T}}, \,\,\,\, \omega_{0,i} = a_{1T,r} - \frac{c_{gT}^2 a_{2T,r}}{4 |a_{2T}|^2}.
\end{equation}
In the framework of linear stability analysis, for $T> T_c$ a negative $\omega_{0,i}$ corresponds to convective instability, while absolute instability happens when $\omega_{0,i}$ is positive. It should be mentioned that the linear AICI is conventionally determined by locating the saddle point in a complex wavenumber plane \citep{Li2019}; in the present analysis $(\beta_0+\alpha_m)$ directly corresponds to such a saddle point.

\subsection{Numerical method}
The highly-accurate numerical simulations of EHD-Poiseuille flows are performed using the open-source computational fluid dynamics solver Nek5000 by Fischer et al. \cite{fischer2008nek5000}, based on the spectral element method (SEM) proposed by Patera \cite{patera1984spectral}. As the current work focuses on the spatiotemporal instability of the EHD-Poiseuille flow, above governing equations (\ref{eq.nlinehd}) and (\ref{eq.linehd}) are numerically solved without any turbulence model. The SEM method combines the accuracy of spectral method with the geometrical flexibility of finite element method. Within each spectral element, the orthogonal basis of Legendre polynomials is used on Gauss-Lobatto-Legendre (GLL) points, which further refine each element and approximate the solution of governing equations as an expansion of Lagrange interpolants. The current spatial discretisation follows the $\mathbb{P}_{N}-\mathbb{P}_{N-2}$ formulation \citep{maday1989spectral} with polynomial order $N=7$. For temporal integration, we use the second-order backward difference scheme implicit for linear terms, coupled to a second-order extrapolation scheme explicit for nonlinear terms. The varying step time is used with Courant number being $0.2$. For the weakly nonlinear stability analysis, we use the spectral collocation method to solve the resultant eigenvalue problems and impose the solvability condition, following Zhang \cite{zhang2016weakly}.

\section{Result and Discussions} \label{sec:results}
\subsection{Grid independence and validation}
In this section, the grid independence of current EHD-Poiseuille flows is first tested by comparing the linear instability criteria $T_c$ and spatiotemporal instability criteria $T_{ca}$ under different element resolutions with parameters $C=10$, $M=10$, $Fe=10^4$, $\mathcal{U}=1$ and domain range $x\in [0,100]$, $y\in [-1,1]$. At $T>T_c$, the electroconvection can take place, and it will transition from convectively to absolutely unstable state when $T>T_{ca}$. Their results are listed in the Table \ref{tab:gridtest}. Based on the polynomial order $N=7$ in the current work, each spectral element further consists of $8\times 8$ GLL nodes. For element size $640\times24$, the total number of grid points should be $4481\times169=757289$ (adjacent elements share the same edge point). From Table \ref{tab:gridtest}, it can be seen that the differences of instability criteria between element sizes $640\times24$ and $800\times28$ are both smaller than $1\%$. Thus, in the following numerical simulations, the spectral element size $640\times24$ is used for streamwise length $L=100$ unless otherwise stated. If other values of $L$ are used, the number of spectral elements along the streamwise direction will be proportionally changed with $L$.

\begin{table}
	\begin{center}
		\setlength{\tabcolsep}{24pt}
		\begin{tabular}{lccc}
			\hline \hline
			Element Size  & $480\times20$ & $640\times24$ & $800\times28$ \\[3pt]
			\hline
			~~$T_c$~~  & ~~$150.16$~~ & ~~$150.77$~~ & ~~$150.81$~~ \\
			~~$T_{ca}$~~  & ~~$265.35$~~ & ~~$265.47$~~ & ~~$265.48$~~ \\
			\hline \hline
		\end{tabular}
		\caption{Grid independence test for EHD-Poiseuille flows by comparing the linear instability criterion $T_c$ and spatiotemporal instability criterion $T_{ca}$ at parameters $C=10$, $M=10$, $Fe=10^4$ and $\mathcal{U}=1$ within the domain range $[0,100]$ in $x$ direction and $[-1,1]$ in $y$ direction. Each spectral element is further refined by GLL points in Nek5000.}
		\label{tab:gridtest}
	\end{center}
\end{table}

\begin{figure}
  \centering
  \includegraphics[width=0.9\textwidth]{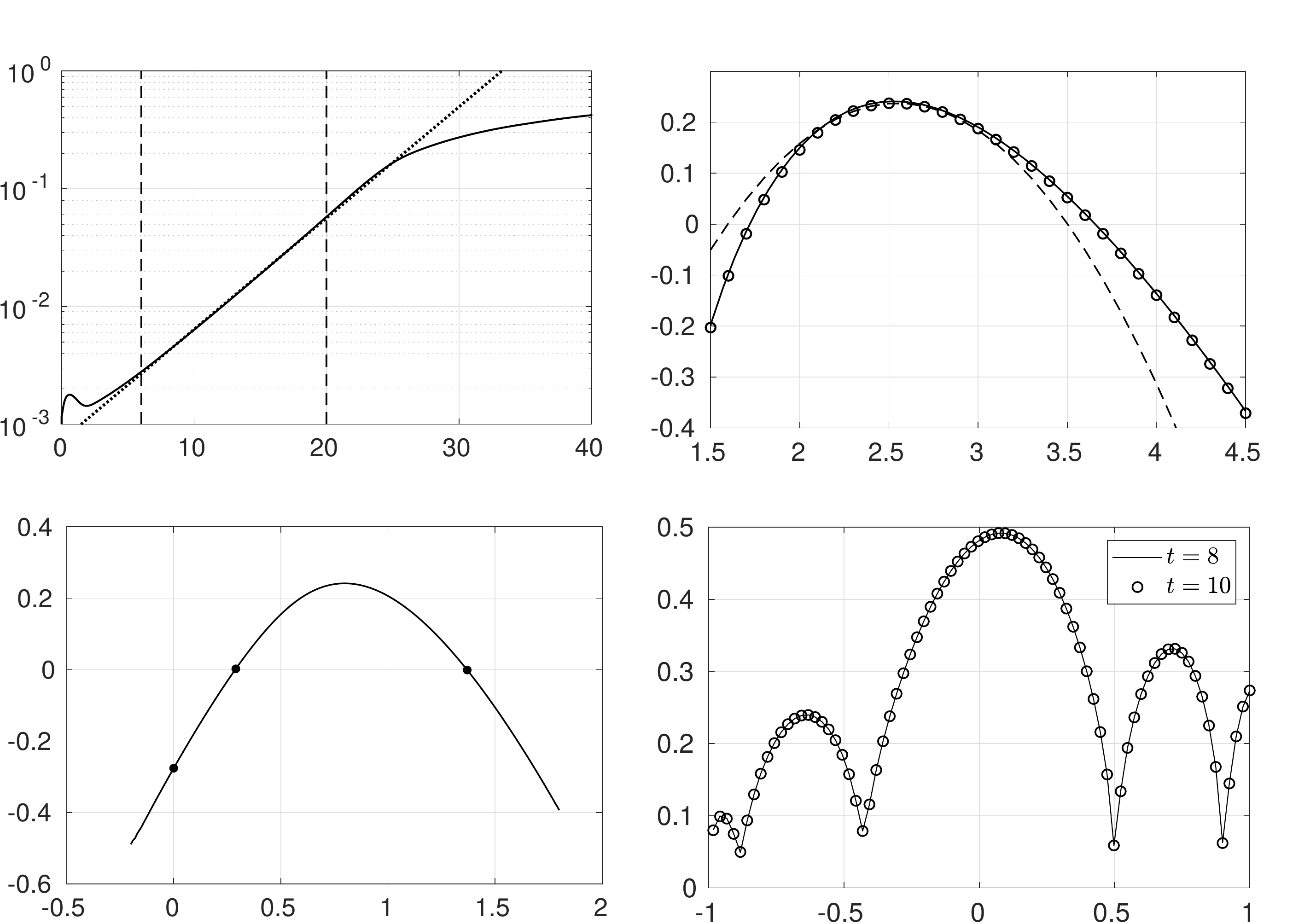}
  \put(-474,291){(a)}
  \put(-345,152){$t$}
  \put(-470,227){$\mathcal{A}$}
  \put(-403,245){linear regime}
  \put(-245,291){(b)}
  \put(-119,152){$\alpha$}
  \put(-240,227){$\omega$}
  \put(-474,135){(c)}
  \put(-346,-5){$v_g$}
  \put(-470,73){$\sigma$}
  \put(-377,80){$v_{-}$}
  \put(-306,80){$v_{+}$}
  \put(-398,47){$v_g=0$}
  \put(-245,135){(d)}
  \put(-116,-5){$y$}
  \put(-240,71){$\tilde{\mathcal{E}}_f$}
  \caption{Temporal and spatiotemporal instabilities of EHD-Poiseuille flows at $C=10$, $M=10$, $Fe=10^4$, $\mathcal{U}=1$ and $T=190$. (a) temporal evolution of global energy amplitude $\mathcal{A}$, the dotted line corresponds to the linearly fitted curve, and two dashed lines indicate the linear regime $t\in[6,20]$; (b) relation between wavenumber $\alpha$ and their leading growth rates $\omega$; the solid line refers to current results, and circles come from the linear stability analysis \citep{zhang2015modal}, and the dashed curve is the result obtained from the parabolic approximation of the dispersion relation as in equation (\ref{eq.linear_GLE1}); (c) relation between the group velocity $v_g=(x-x_0)/t$ and $\omega$; (d) invariant eigenvectors $\tilde{\mathcal{E}}_f$ at $\alpha=2.513$ (corresponds to the maximum growth rate in panel b) at two different instants.}
 \label{fig.validateT190}
\end{figure}
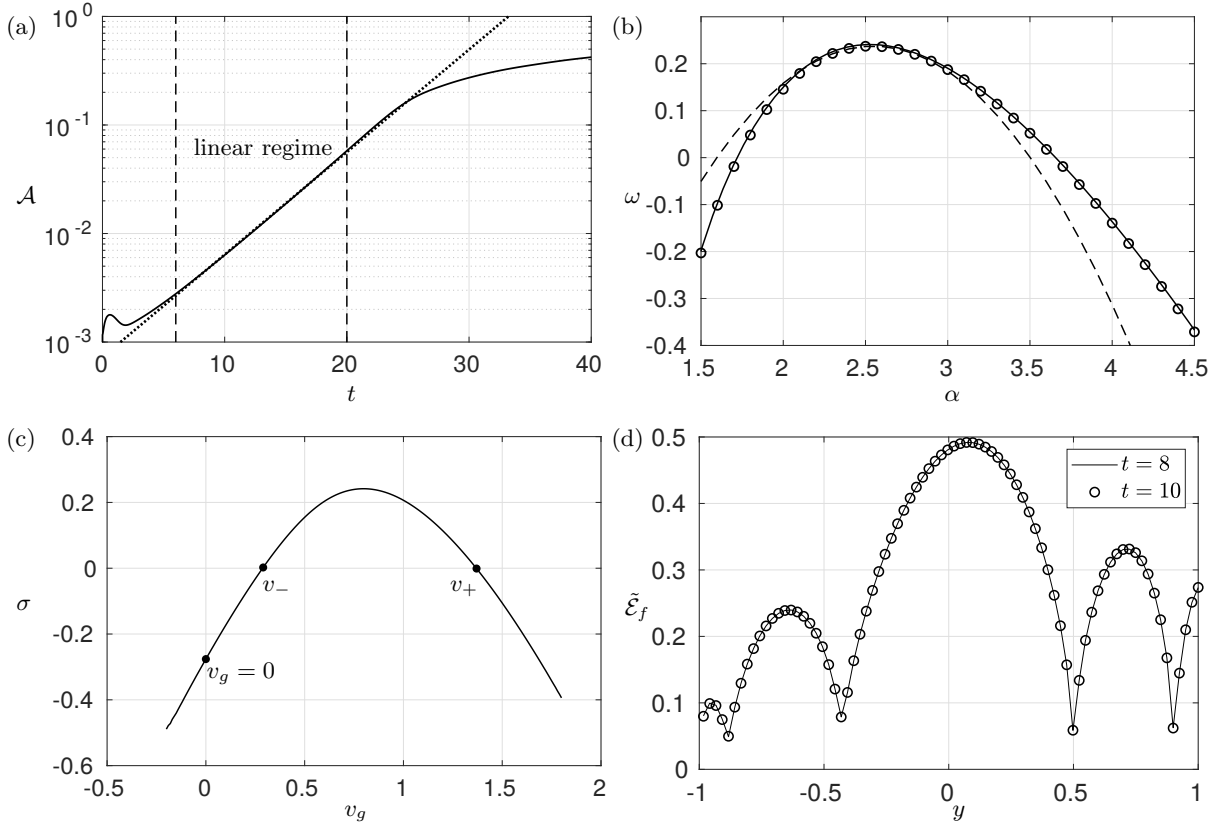

The accuracy of numerical simulations for electroconvective flows within the framework of Nek5000 has been verified in our former work \cite{feng2021deterministic}. Here, the capacity of the flow solver to study the temporal and spatiotemporal instability is tested by analysing the snapshots of the EHD-Poiseuille flows. To do so, we set the spatial domain range to $x\in[0,100]$ and $y\in[-1,1]$, and use the control parameters $C=10$, $M=10$, $Fe=10^4$, $\mathcal{U}=1$ and $T=190$. Nonlinear governing equations (\ref{eq.nlinehd}) are evolved to obtain snapshots of flow fields within the linear regime (about $t\in[6,20]$) as shown in Fig. \ref{fig.validateT190}(a). The results of temporal instability analysis are shown in panel b, where the leading growth rates (i.e. solid line) at different wavenumbers are calculated from the data of numerical simulations and they agree well with those from the local linear stability analysis \citep{zhang2015modal} (circles). The maximum growth rate is $\omega=0.2409$ at wavenumber $\alpha=2.513$. For the spatiotemporal instability analysis in panel c, the maximum growth rate $\omega=0.2416$ is obtained at group velocity $v_g=0.800$, which is consistent with above growth rate $\omega=0.2409$ with a relative difference of $0.290\%$. At $v_g=0$, the growth rate is negative $\sigma=-0.276$ which indicates a linearly convective type of instability for the EHD-Poiseuille flow at $T=190$. The leading velocity $v_{+}=1.370$ and the trailing velocity $v_{-}=0.290$ are also shown in panel c. They have the same sign (i.e. positive) and the ray $v_g=0$ lies outside the range of $[v_{-},v_{+}]$. Finally, the eigenvector $\mathcal{E}_f$ of wavenumber $\alpha=2.513$ at $t=8$ and $t=10$ coincide with each other as shown in panel d, which means that they are proportional to each other (as they are at the asymptotically large time) and the current calculated growth rates are faithful.

\begin{figure}
  \centering
  \includegraphics[width=0.5\textwidth]{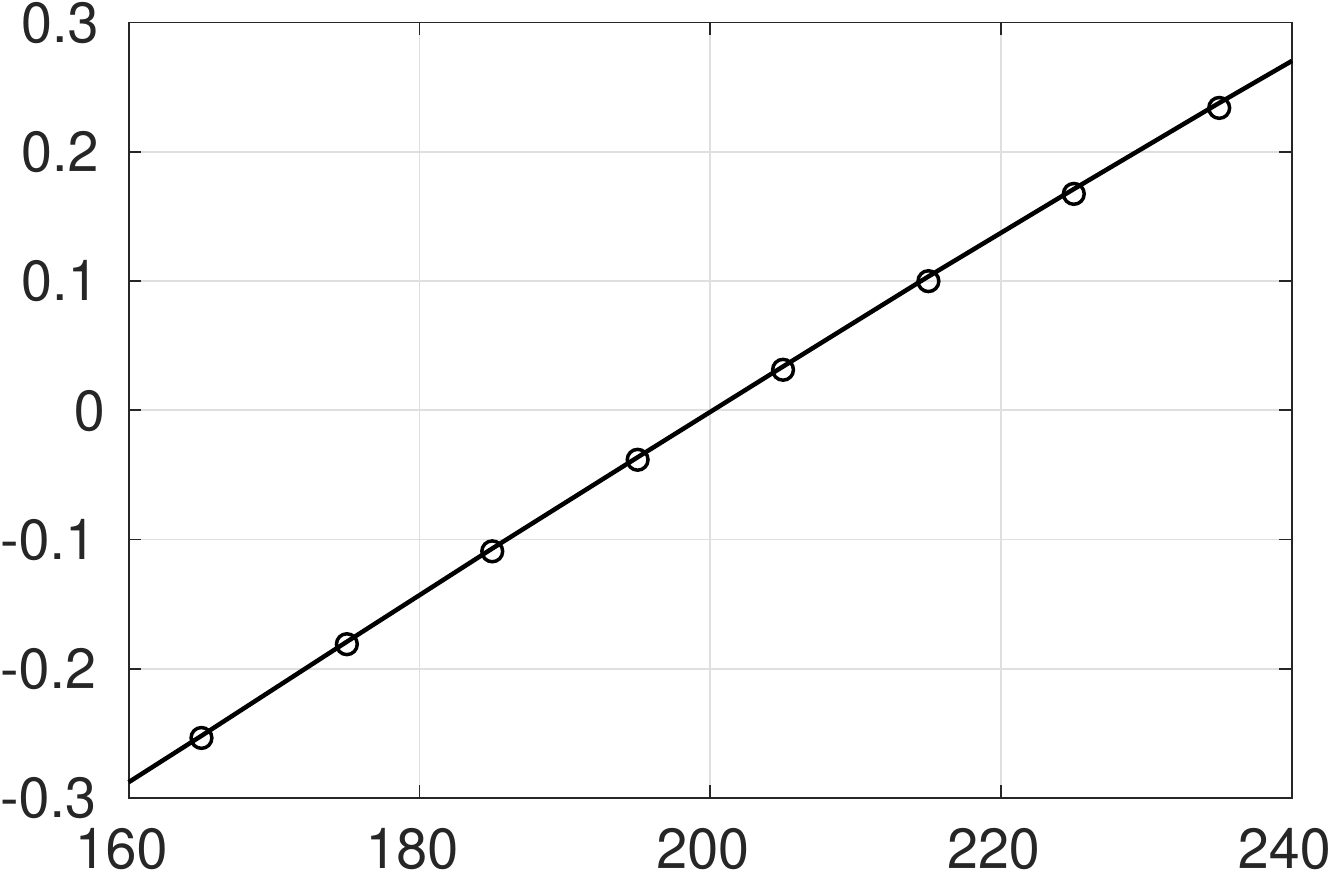}
  \put(-123,-10){$T$}
  \put(-265,86){$\sigma$}
  \caption{Variation of absolute growth rate with the electric Rayleigh number for EHD-Poiseuille flows at $C=50$, $M=50$, $Fe=10^4$ and $\mathcal{U}=1$. The circular symbols refer to current results, and solid line comes from the work of \cite{Li2019}. $\sigma<0$: linearly convective instability; $\sigma>0$: linearly absolute instability; $\sigma=0$: linearly neutral, $T_{ca}=200.89$.}
 \label{fig.valiLi}
\end{figure}

In addition, the absolute growth rates $\sigma(v_g=0)$ are computed for discriminating the transition from linearly convective instability to absolute instability, and they are compared with the linear results from Li et al. \cite{Li2019}. The parameters are $C=50$, $M=50$, $Fe=10^4$ and $\mathcal{U}=1$. As the injection strength increases to $C=50$, the mesh resolution in vertical direction is further refined to have $36$ spectral elements. The current results are shown in Fig. \ref{fig.valiLi}. It can be found that the absolute growth rates and criterion of absolute instability (i.e. $T_{ca}=200.89$ in the current work) agree well with the linear results in Ref. \citep{Li2019} (whose $T_{ca}=200.37$), which verifies the accuracy of the current algorithm for performing the spatiotemporal instability analysis of the EHD-Poiseuille flows.

\subsection{Weakly nonlinear stability analysis}
In this section, the characteristics of transition from hydrostatic to electroconvective state around the linear instability criterion $T_c$ are first studied. In order to evaluate Landau coefficients from the data of numerical simulations, we assume the following Stuart-Landau equation
\begin{equation}
\frac{d\mathcal{A}}{dt} = c_1\mathcal{A} + c_3\mathcal{A}|\mathcal{A}|^2, \,\,\, \frac{d \ln \mathcal{A}}{dt} = c_1 + c_3|\mathcal{A}|^2.
\end{equation}
After obtaining the time series of global energy amplitude $\mathcal{A}(t)$ in Eq. \ref{eq.gamplitude} from numerical simulations, $\frac{d \ln \mathcal{A}}{dt}$ can be plotted as a function of $|\mathcal{A}|^2$. Then $c_1$ can be approximated by the vertical intercept and $c_3$ by the slope. In terms of the coefficients obtained from the weakly nonlinear analysis, $c_1$ should be compared with $\mu$ and $\sigma$ and $c_3$ should be compared with $a_3$ and $K$.

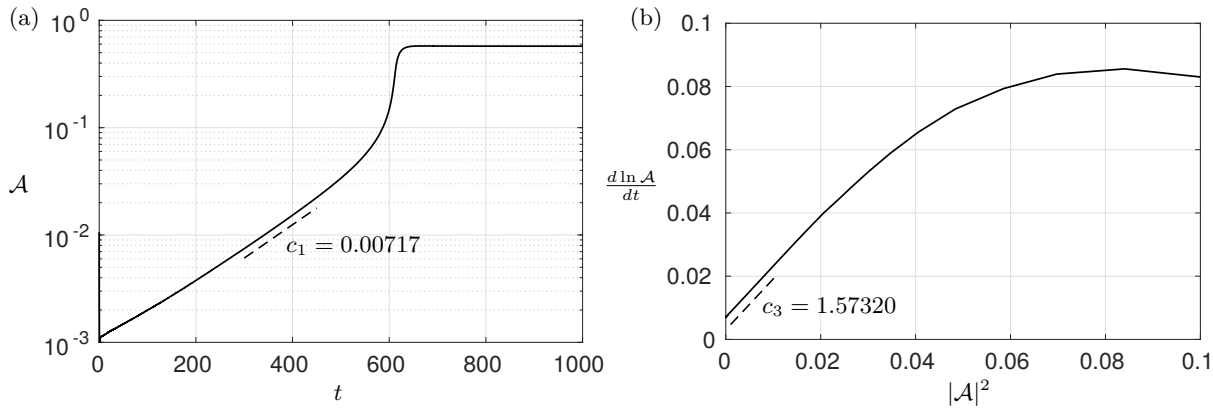
\begin{figure}
  \centering
  \includegraphics[width=0.9\textwidth]{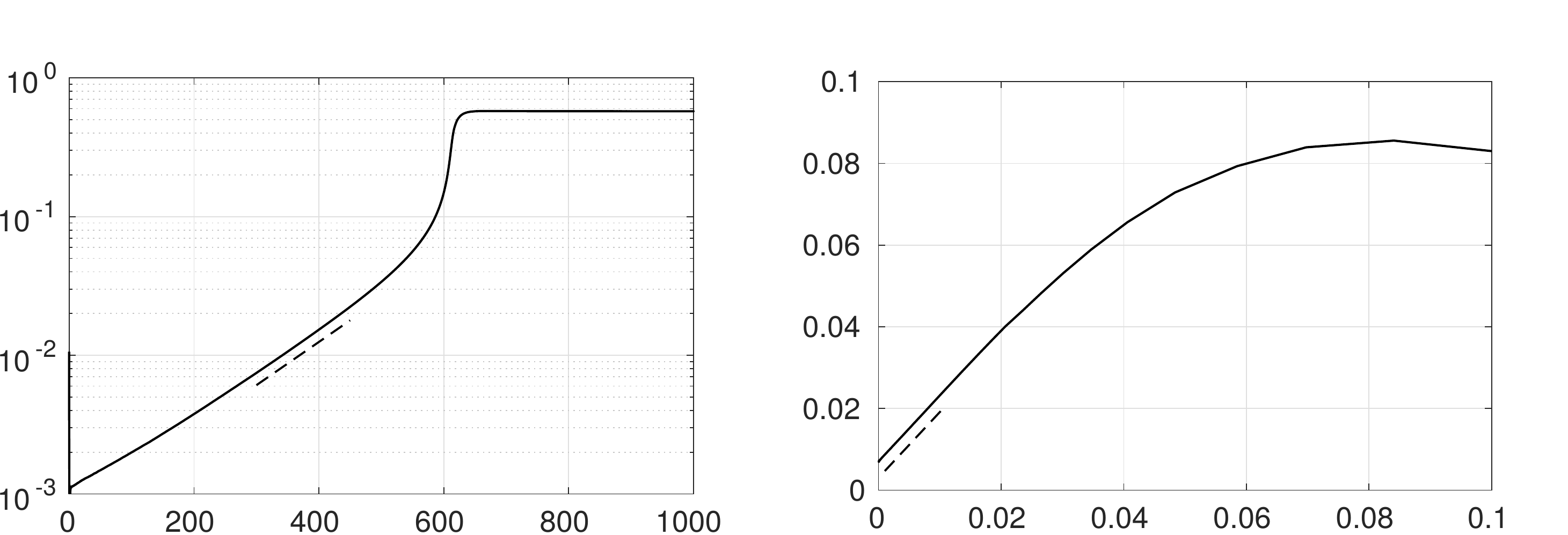}
  \put(-473,132){(a)}
  \put(-473,69){$\mathcal{A}$}
  \put(-350,-10){$t$}
  \put(-368,46){$c_1=0.00717$}
  \put(-238,132){(b)}
  \put(-248,69){$\frac{d\ln \mathcal{A}}{dt}$}
  \put(-118,-10){$|\mathcal{A}|^2$}
  \put(-188,23){$c_3=1.57320$}
  \caption{The first transition of EHD-Poiseuille flows characterized by the global energy amplitude $\mathcal{A}$ at $C=10$, $M=10$, $Fe=10^4$, $\mathcal{U}=1$ and $T=152$. (a) temporal evolution of $\mathcal{A}$; (b) relation between $d(\ln \mathcal{A})/dt$ and $|\mathcal{A}|^2$.}
 \label{fig.dlnAdt_T152}
\end{figure}

With the parameters $C=10$, $M=10$, $Fe=10^4$, $\mathcal{U}=1$ and $T=152$ in a computational domain $x\in[0,2.456]$ and $y\in[-1,1]$, the results of the growth rate $c_1$ and the Landau coefficient $c_3$ can be computed as $c_1=0.00717$ and $c_3=1.57320$, as shown in Fig. \ref{fig.dlnAdt_T152}, by performing the linear-least-square fitting of the data in panels a and b (we will compare these results with those in the weakly nonlinear stability theory to be followed). Around $|\mathcal{A}|^2 = 0$ in panel b, the perturbation amplitude is weak and the system stays within the linear regime. Thus, the $y$-intercept corresponds the growth rate $c_1$ obtained in panel a. In the current case, the Landau coefficient $c_3$ is found to be positve at $T=152$ close to the linear instability criterion $T_c=150.77$, which indicates that the transition from hydrostatic to electroconvective state is subcritical in the EHD-Poiseuille flows  \cite{zhang2016weakly,guan2019numerical} at the aforementioned parameters. From the viewpoint of temporal evolution of $\mathcal{A}$ in panel a, the deviation from the linear growth after $t=580$ is associated with an increased growth rate, which indicates that the higher-order terms destabilises the flow (around the linear critical condition), consistent with the subcritical bifurcation.

\begin{figure}
  \centering
  \includegraphics[width=0.9\textwidth]{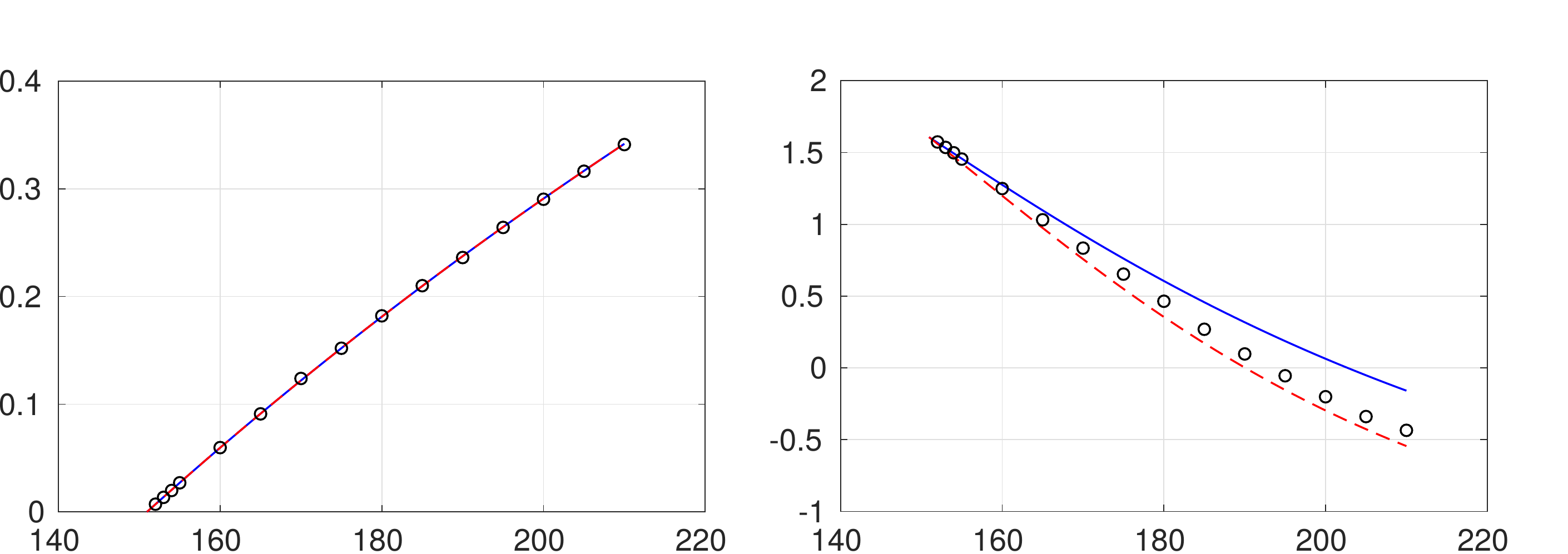}
  \put(-475,136){(a)}
  \put(-473,73){$c_1$}
  \put(-351,-10){$T$}
  \put(-244,136){(b)}
  \put(-243,73){$c_3$}
  \put(-121,-10){$T$}
  \caption{Comparison of the linear growth rate and Landau coefficient between the data of numerical simulations (circular symbols) and those from the multiple-scale method (solid blue line) and amplitude expansion method (red dashed line) at $C=10$, $M=10$, $Fe=10^4$, $\mathcal{U}=1$. (a) growth rate $c_1$; (b) Landau coefficient $c_3$.}
 \label{fig.constM10}
\end{figure}

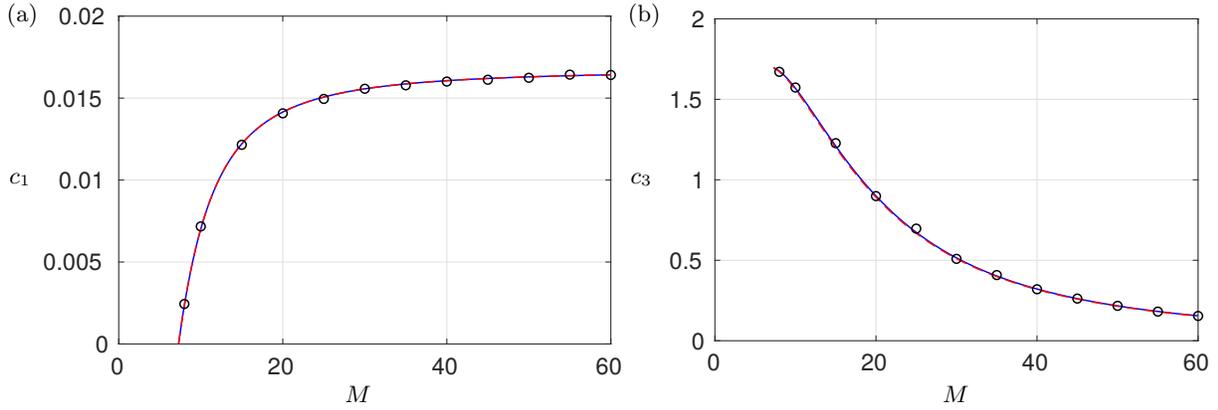
\begin{figure}
  \centering
  \includegraphics[width=0.9\textwidth]{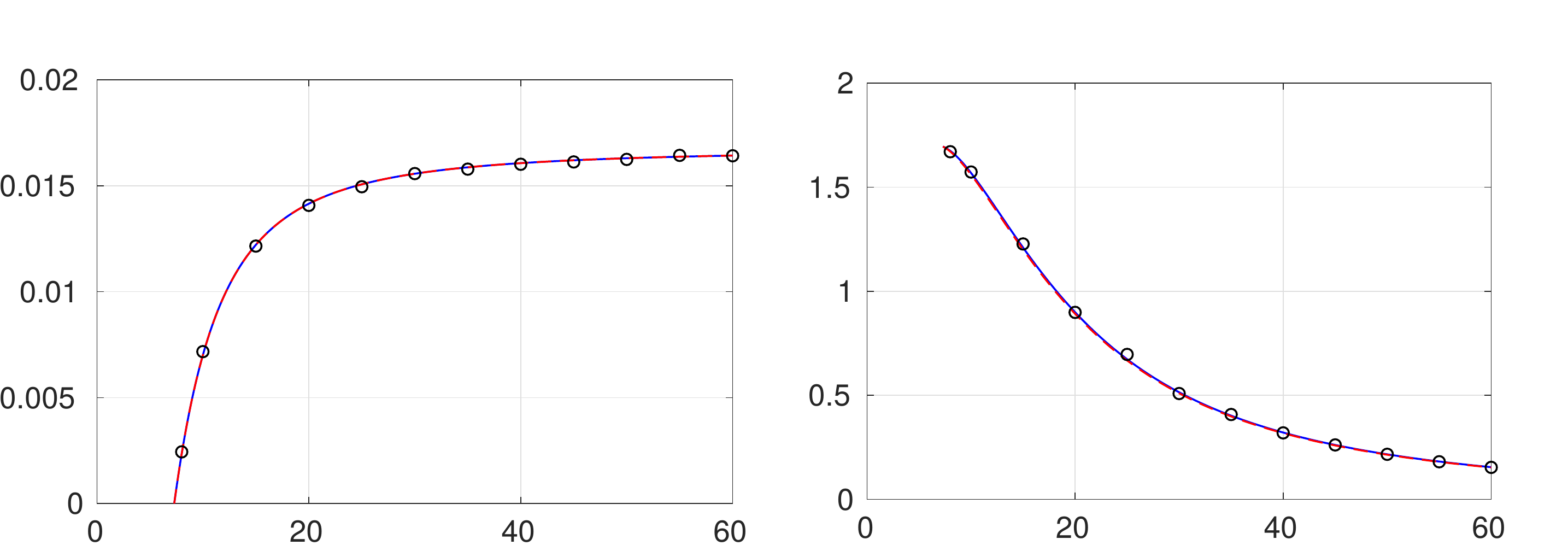}
  \put(-473,134){(a)}
  \put(-472,73){$c_1$}
  \put(-345,-10){$M$}
  \put(-238,134){(b)}
  \put(-237,73){$c_3$}
  \put(-119,-10){$M$}
  \caption{Comparison of the linear growth rate and Landau coefficient between current results (circular symbols) and those from the multiple-scale method (solid blue line) and amplitude expansion method (red dashed line) at $C=10$, $T=152$, $Fe=10^4$, $\mathcal{U}=1$. (a) growth rate $c_1$; (b) Landau coefficient $c_3$.}
 \label{fig.constT152}
\end{figure}

In the above paragraph, the linear growth rate and the Landau coefficient are approximately obtained from the time series of global energy amplitude in numerical simulations. On the other hand, starting from the governing equations (\ref{eq.nlinehd}) of EHD-Poiseuille flows, the GLE can be explicitly derived by perturbation methods. We will consider both the multiple-scale method and amplitude expansion method and compare their results with the linear growth rate and the Landau coefficient obtained from the data of numerical simulations.
As shown in Fig. \ref{fig.constM10}, the relation between $c_1$, $c_3$ and $T$ is investigated with other parameters being invariant. The linear growth rates $c_1$ obtained from three methods are visually the same to each other. For Landau coefficient $c_3$ in panel b, around the linear instability criterion $T_c=150.77$, $c_3$ predicted from three methods agree well with each other, which indicates the equivalence of the two perturbation methods around the linear instability criterion. However, as the increase of electric Rayleigh number from $T_c$, the value of $c_3$ obtained from the data of numerical simulations gets closer to that of the amplitude expansion method. This clearly confirms that the validity of amplitude expansion method is wider than multiple-scale method when the governing parameter $T$ is away from the linear critical value $T_c$. Except for the effect of electric Rayleigh number on $c_1$ and $c_3$, the mobility ratio $M$ is also changed to display the dependence of $c_1$ and $c_3$ on it, as shown in Fig. \ref{fig.constT152}. The electric Rayleigh number is fixed at $T=152$ which is very close to the linear instability criterion $T_c=150.77$. Around this $T$, the multiple-scale method and amplitude expansion method are approximately equivalent. Thus, the results of $c_1$ and $c_3$ obtained from three methods are consistent with each other. The reason may be that changing $M$ will not change the linear critical condition \cite{zhang2015modal} and will not drive the flow away from the linear critical condition (note that the $T$ used is close to $T_c$).

\begin{figure}
  \centering
  \includegraphics[width=0.5\textwidth]{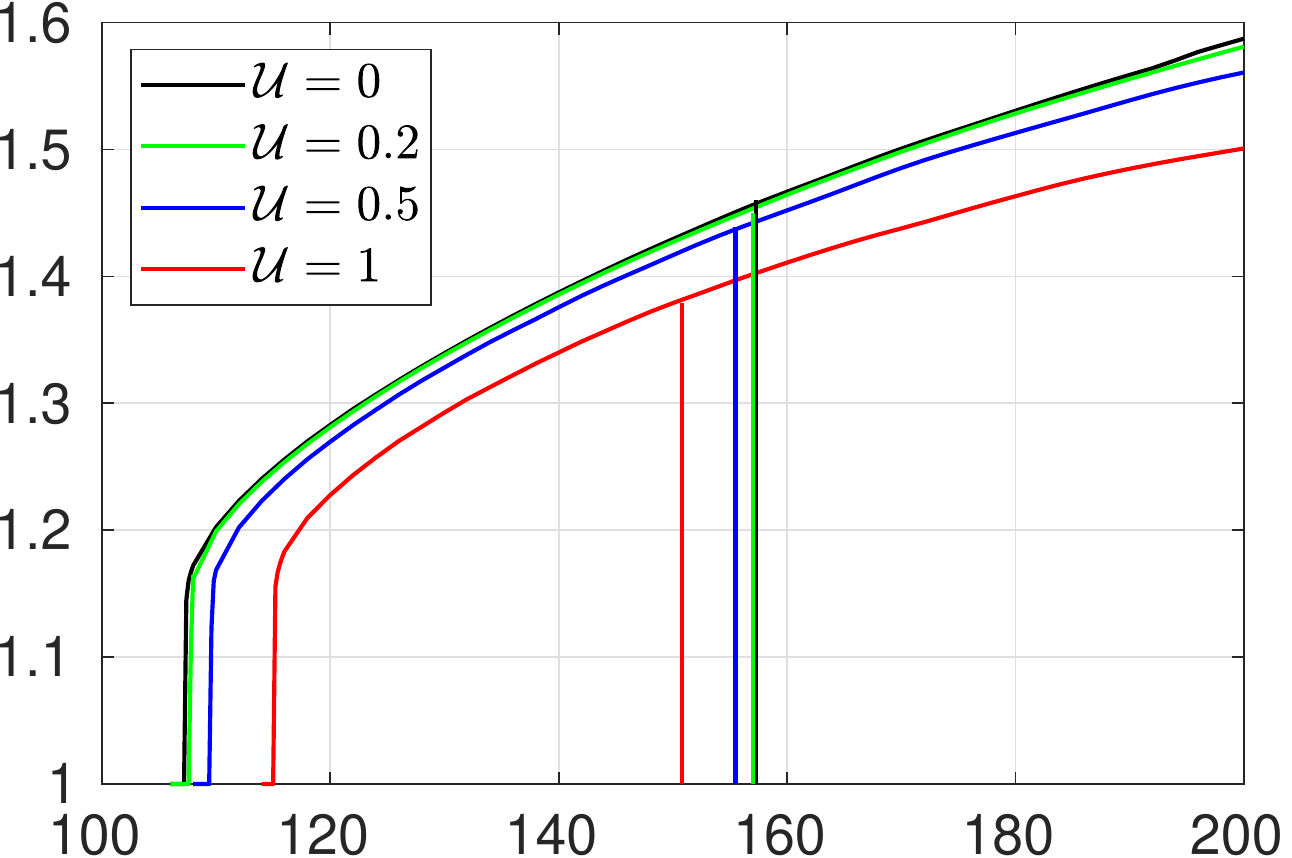}
  \put(-272,87){$Ne$}
  \put(-124,-10){$T$}
  \put(-130,19){$T_c$}
  \put(-199,19){$T_f$}
  \caption{Hysteresis loops of EHD-Poiseuille flows at $C=10$, $M=10$, $Fe=10^4$ and different $\mathcal{U}$. (1) Black solid line: $\mathcal{U}=0$, $T_c=157.27$ and $T_f=107.4$; (2) Green solid line: $\mathcal{U}=0.2$, $T_c=156.98$ and $T_f=107.8$; (3) Blue solid line: $\mathcal{U}=0.5$, $T_c=155.48$ and $T_f=109.6$; (4) Red solid line: $\mathcal{U}=1$, $T_c=150.77$ and $T_f=115.2$.}
 \label{fig.hystloop_varU}
\end{figure}
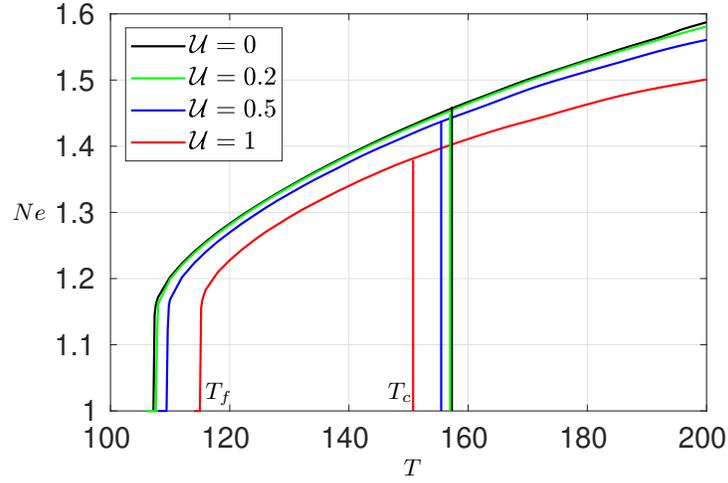

One salient feature of the subcritical transition is the existence of a hysteresis loop in the bifurcation diagram, which has been studied and discussed in the electroconvection without through-flows (e.g., in our previous work \cite{feng2021deterministic}). When exposed to the Poiseuille flow, the linear instability criterion $T_c$ and finite-amplitude criterion $T_f$ forming the hysteresis loop can present different results compared with the case without through-flows. With parameters $C=10$, $M=10$ and $Fe=10^4$, the hysteresis loops at $\mathcal{U}\in[0,0.2,0.5,1]$ are computed as shown in Fig. \ref{fig.hystloop_varU}. The electric Nusselt number $Ne$ within the range $T\in[T_f,T_c]$ is obtained by gradually decreasing $T$. When the through-flow is intensified, it can be seen that the electric transferring efficiency is reduced (i.e. $Ne$ becomes smaller) as the vertical motions are weakened by the through-flow \cite{guan2019numerical}. The results of linear instability criteria $T_c$ agree well with former results \citep{zhang2015modal} and $T_f$ increases. Furthermore, as the increase of $\mathcal{U}$ from $0$ to $1$, the range between $T_f$ and $T_c$ decreases, which indicates that the subcritical nature of the pure electroconvective flows can be suppressed by the Poiseuille flow.

\begin{figure}
  \centering
  \includegraphics[width=0.6\textwidth]{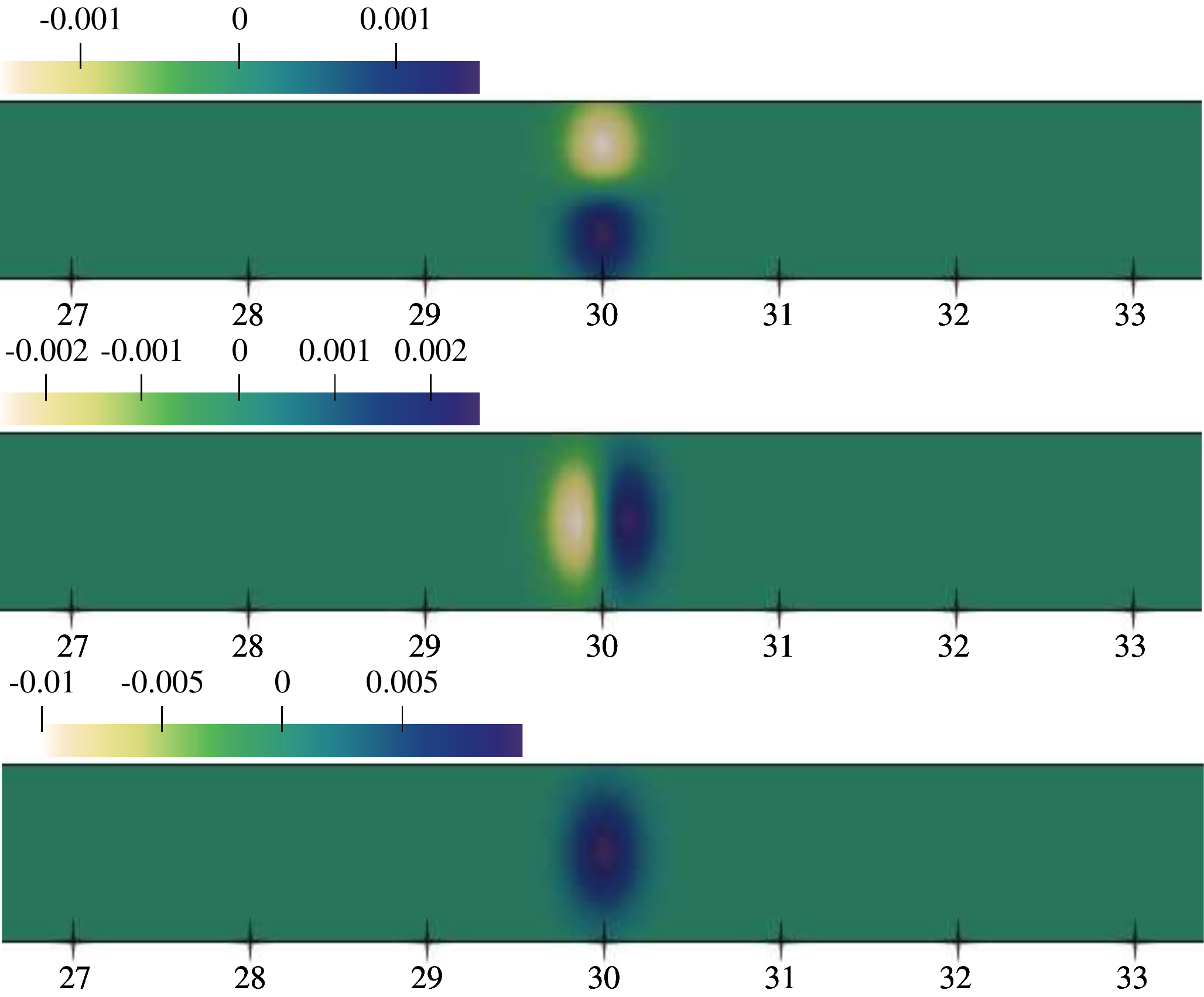}
  \put(-325,240){(a)}
  \put(-314,203){$y$}
  \put(-325,152){(b)}
  \put(-314,119){$y$}
  \put(-325,72){(c)}
  \put(-314,34){$y$}
  \put(-154,-5){$x$}
  \caption{Distribution of the initially localised impulse defined by equations (\ref{eq.impulse}) in the EHD-Poiseuille flow. Only part of the flow field is presented here for a clear visualisation as the impulse is located around the central position $x_0=30$ and $y_0=0$. (a) streamwise velocity; (b) normal velocity; (c) charge density.}
 \label{fig.initImpulse}
\end{figure}
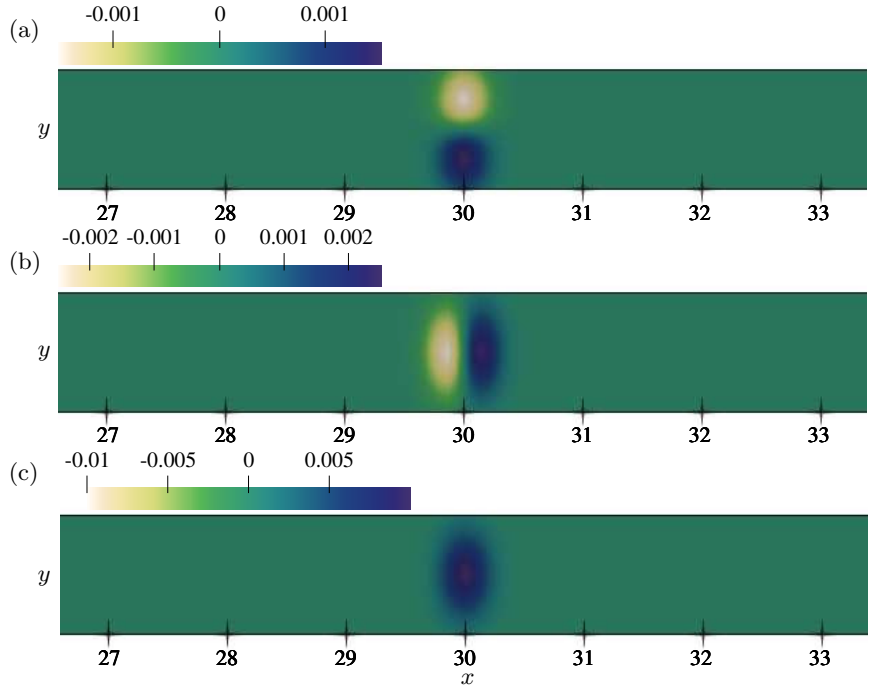

\subsection{Spatiotemporal instability analysis}
Next, the convective and absolute instabilities of EHD-Poiseuille flows are analyzed by studying both linear and nonlinear responses of the flow field to an initially localized impulse. For the profile of the impulse, we follow that in Ref. \citep{delbende1998nonlinear} 
\begin{subequations} \label{eq.impulse}
\begin{align} 
  u(x,y,t=0) &= -A_m(y-y_0) \exp \left[ -\left( \frac{(x-x_0)^2}{2\sigma_x^2} + \frac{(y-y_0)^2}{2\sigma_y^2} \right) \right], \\
  v(x,y,t=0) &= A_m(x-x_0) \frac{\sigma_y^2}{\sigma_x^2} \exp \left[ -\left( \frac{(x-x_0)^2}{2\sigma_x^2} + \frac{(y-y_0)^2}{2\sigma_y^2} \right) \right], \\
  q(x,y,t=0) &= A_m\exp \left[ -\left( \frac{(x-x_0)^2}{2\sigma_x^2} + \frac{(y-y_0)^2}{2\sigma_y^2} \right) \right],
\end{align}
\end{subequations}
where $(x_0,y_0)$ refers to the central location of the impulse, and the extent $\sigma_x=0.3$ along $x$-axis and $\sigma_y=0.2$ along $y$-axis. The initial perturbation amplitude is set to be $A_m=10^{-3}$. The envelope of this impulsed disturbance takes the Gaussian form, and two initial velocity components $u$ and $v$ satisfy the requirement of continuity equations (\ref{eq.linehd}a). This localized impulse initially added into the EHD-Poiseuille flow is shown in Fig. \ref{fig.initImpulse} for reference.

\begin{figure}
  \centering
  \includegraphics[width=0.92\textwidth]{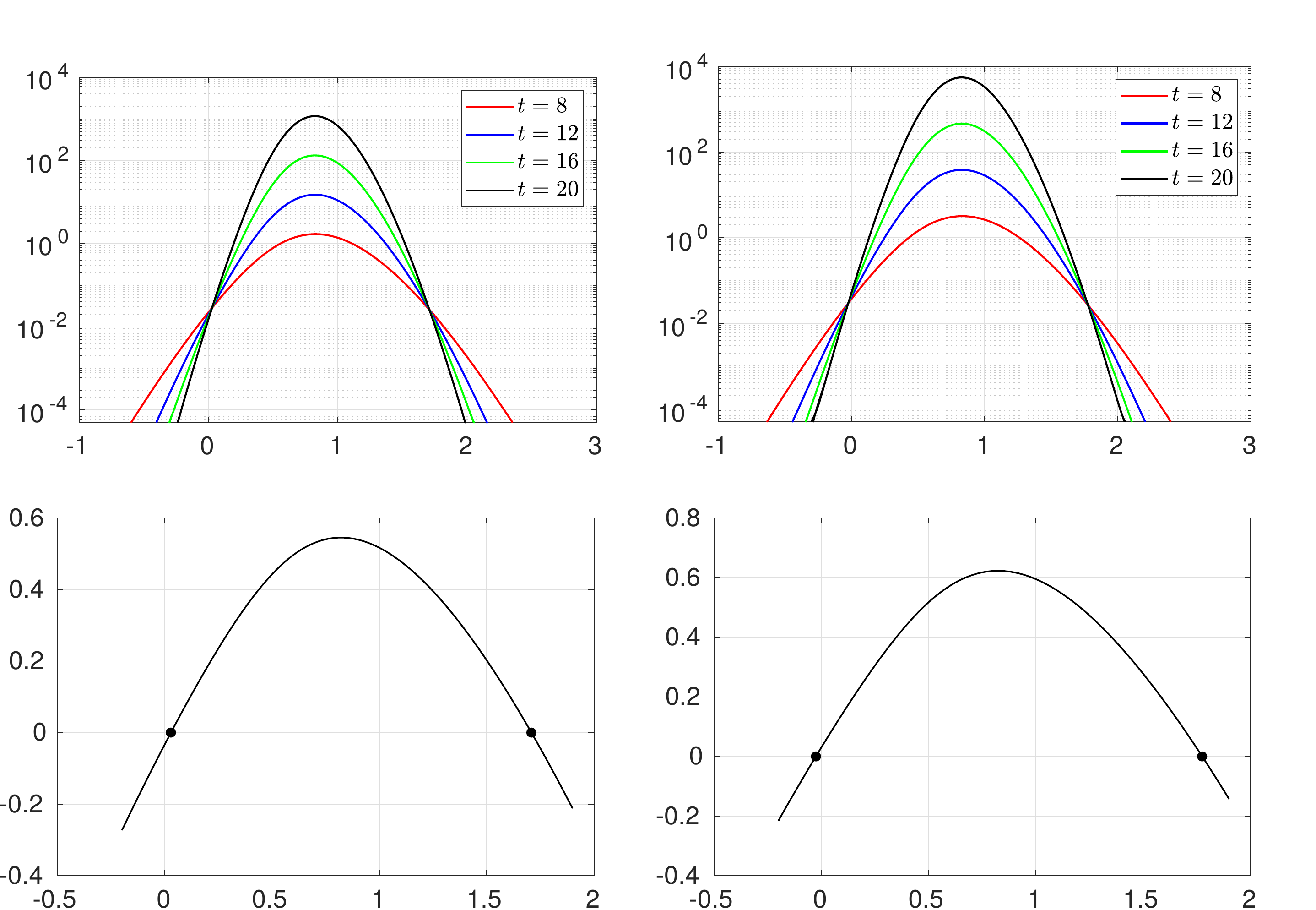}
  \put(-482,295){(a)}
  \put(-356,155){$v_g$}
  \put(-485,230){$t^{\frac{1}{2}}\hat{\tilde{\mathcal{E}}}$}
  \put(-438,285){$T=255$}
  \put(-248,295){(b)}
  \put(-122,155){$v_g$}
  \put(-252,230){$t^{\frac{1}{2}}\hat{\tilde{\mathcal{E}}}$}
  \put(-208,288){$T=275$}
  \put(-482,135){(c)}
  \put(-356,-8){$v_g$}
  \put(-485,73){$\sigma$}
  \put(-444,122){$T=255$}
  \put(-408,55){$v_{-}=0.028$}
  \put(-325,55){$v_{+}=1.708$}
  \put(-248,135){(d)}
  \put(-122,-8){$v_g$}
  \put(-252,73){$\sigma$}
  \put(-212,122){$T=275$}
  \put(-177,46){$v_{-}=-0.025$}
  \put(-86,46){$v_{+}=1.775$}
  \caption{Variation of $t^{\frac{1}{2}}\hat{\tilde{\mathcal{E}}}$ and $\sigma$ with the group velocity $v_g$ for EHD-Poiseuille flows at $C=10$, $M=10$, $Fe=10^4$ and $\mathcal{U}=1$  for two values of $T$. (a-b) Temporal evolution of $t^{\frac{1}{2}}\hat{\tilde{\mathcal{E}}}$; (c-d) spatiotemporal growth rate $\sigma$ as a function of $v_g$.}
 \label{fig.tamp_vg}
\end{figure}
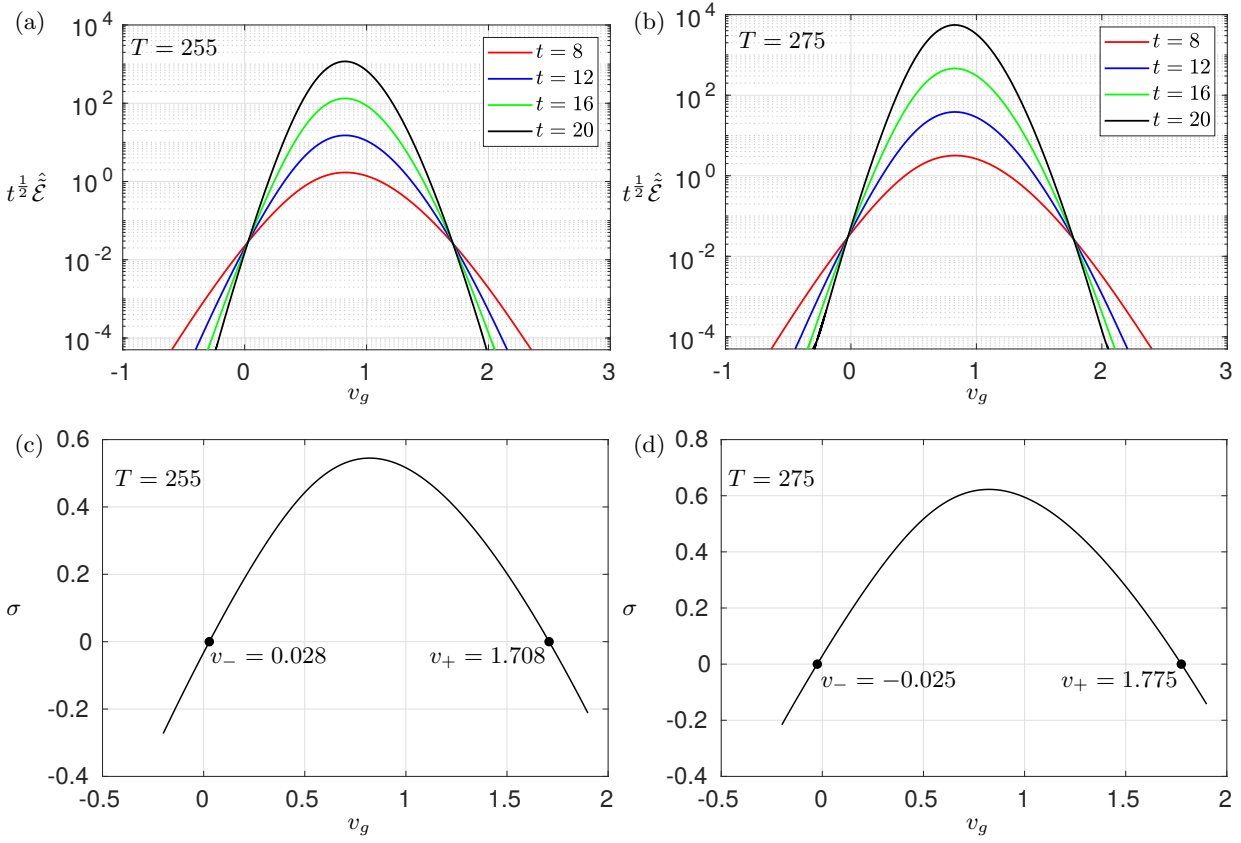

After the onset of electroconvective flows (i.e. $T>T_c$), under the effects of the electric field and the through-flow, the initial impulse in the flow field can develop both temporally and spatially. To study the linear and nonlinear responses of such impulse under periodic streamwise boundaries and to avoid contamination, the spatial range along the streamwise direction is extended to $x\in[0,100]$ to prevent the re-entrance of this impulse from the outlet to inlet of this spatial domain. The parameters $C=10$, $M=10$ and $Fe=10^4$ are kept constant in this section. 


\subsubsection{Linear impulse response}\label{sec:linear_AICI}

The transition from linearly convective instability to linearly absolute instability can be discriminated by the absolute growth rate $\sigma$ of the ray $v_g=0$, which varies almost linearly with the driving parameter $T$ (see Fig. \ref{fig.valiLi}). As defined in equation (\ref{eq.stasym}), the spatiotemporal growth rate $\sigma$ can be determined by the exponentially temporal variation of amplitude $t^{\frac{1}{2}}\hat{\tilde{\mathcal{E}}}$ at a given group velocity $v_g$. Their relations are presented in Fig. \ref{fig.tamp_vg}(a-b) for two values of $T$. It can be seen that the crossings of these amplitude curves define two group velocities (i.e. $v_{-}$ and $v_{+}$) where the amplitude $t^{\frac{1}{2}}\hat{\tilde{\mathcal{E}}}$ neither grows nor decays. Within these two limiting edges $v_g\in[v_{-},v_{+}]$, the wavepacket always grows. Based on the equation (\ref{eq.sigma}), the spatiotemporal growth rates can be calculated at different group velocities as shown in panels c and d. At $T=255$, the amplitude curve at $v_g=0$ in panel a decays, which indicates that $\sigma(v_g=0)$ is negative and the flow is convectively unstable (or $v_{-}$ and $v_{+}$ share the same direction). As the electric Rayleigh number is increased to $T=275$, the amplitude curve in panel b grows at $v_g=0$ and $\sigma(v_g=0)>0$, which indicates that the flow becomes absolutely unstable (or the signs of $v_{-}$ and $v_{+}$ are different). 

\begin{table}
	\begin{center}
		\setlength{\tabcolsep}{8pt}
		\begin{tabular}{clcccc}
			\hline \hline
			Results from&$T$  & $150.95\,\ (T_c)$ & $190$ & $250$ & $300$ \\ 
			\hline
			&$c_{g}$  &$0.7813$ &$0.8029$ &$0.8214$ &$0.8303$\\
			multiple-scale&$a_{1,r}(T-T_c)$  & $0$ &$0.2139$ &$0.4096$ &$0.5000$\\
			expansion&$a_{2}$  &$0.2613-0.0177i$ &$0.2614-0.0252i$ &$0.2622-0.0275i$ &$0.2614-0.0274i$\\
			&$\omega_{0M,i}$  &$-0.5814$ &$-0.3969$ &$-0.2267$ &$-0.1523$\\ \hline
			&$c_{gA}$  & $0.7813$ & $0.8029$ & $0.8214$ & $0.8303$ \\
			amplitude&$\sigma_{r}$  & $0.0000$ & $0.2374$ & $0.5230$ & $0.7093$ \\
			expansion&$a_{2A}$  & $0.2613-0.0177i$ & $0.2614-0.0252i$ & $0.2622-0.0275i$ & $0.2613-0.0274i$ \\
			&$\omega_{0A,i}$  &$-0.5814$ &$-0.3734$ &$-0.1133$ &$0.0570$\\ \hline			
			&$c_{gT}$  & $0.7812$ & $0.8028$ & $0.8214$ & $0.8302$ \\
			dispersion&$a_{1T,r}$  & $-0.0000$ & $0.2374$ & $0.5230$ & $0.7092$ \\
			relation&$a_{2T}$  & $0.2615-0.0176i$ & $0.2616-0.0251i$ & $0.2624-0.0275i$ & $0.2614-0.0273i$ \\
			&$\omega_{0,i}$  &$-0.5809$ &$-0.3729$ &$-0.1129$ &$0.0573$\\ \hline
			Relative&$|(\omega_{0M,i}-\omega_{0,i})/\omega_{0,i}|$  &$0.09\%$ &$6.44\%$ &$100.80\%$ &$365.79\%$\\	
			errors&$|(\omega_{0A,i}-\omega_{0,i})/\omega_{0,i}|$  &$0.09\%$ &$0.13\%$ &$0.35\%$ &$0.52\%$\\		
			\hline \hline
		\end{tabular}
		\caption{Coefficients in the linearised GLE for EHD-Poiseuille flows at different $T$ and parameters $C=10$, $M=10$, $Fe=10^4$ and $\mathcal{U}=1$. The coefficients $c_{g}$, $a_{1,r}$ and $a_{2}$ are calculated based on the multiple-scale expansion using equations (\ref{eq.cg}) and (\ref{eq.Landau-coefficients}); the corresponding absolute growth rate is $\omega_{0M,i} = a_{1,r}(T-T_c) - {c_{g}^2 a_{2,r}}/({4 |a_{2}|^2})$. The coefficients $c_{gA}$, $\sigma_{r}$ and $a_{2A}$ are calculated based on the amplitude expansion using equations (\ref{eq.amplitude_equations}a) and (\ref{eq.cgA_a2A}); the corresponding absolute growth rate is $\omega_{0A,i} = \sigma_{r} - {c_{gA}^2 a_{2A,r}}/({4 |a_{2A}|^2})$. The coefficients $c_{gT}$, $a_{1T,r}$ and $a_{2T}$ are directly evaluated from the dispersion relation using equation (\ref{eq.coeff_dispersion}), and the corresponding absolute growth rate is calculated using equation (\ref{eq.abs_growthrate}).The last two rows show the relative errors between the absolute growth rate directly obtained from the dispersion relation and those obtained using the two expansion methods.} 
		\label{tab:comparison_cga1a2}
	\end{center}
\end{table}

As mentioned in section \ref{sec:AICI_GLE}, the absolute growth rate $\omega_{0,i}$ in equation (\ref{eq.abs_growthrate}), which shares the same meaning as $\sigma(v_g=0)$ defined in equation (\ref{eq.sigma}), can also be used to determine the AI/CI property (here we use different notations for the absolute growth rate to differentiate it in different methods, i.e., $\omega_{0,i}$ in the weakly nonlinear analysis and $\sigma(v_g=0)$ in the post-processing analysis of snapshots from numerical simulations). To calculate $\omega_{0,i}$, the three coefficients in the linearised GLE should be evaluated first. As we have explained above, there are three methods for doing this: the first is to calculate the coefficients $c_{g}$, $a_{1}$ and $a_{2}$ in equation (\ref{eq.linear_GLE}a) based on the formulae (\ref{eq.cg}) and (\ref{eq.Landau-coefficients}); the second is to calculate $c_{gA}$, $\sigma$ and $a_{2A}$ in equation (\ref{eq.linear_GLE}b) according to the formulae (\ref{eq.amplitude_equations}a) and (\ref{eq.cgA_a2A}); and the third is to directly use the dispersion relation given in equation (\ref{eq.coeff_dispersion}) which corresponds to the linear eigenvalue problem (\ref{eq.epsilons}a). In order to quantify the differences between these three methods, Table \ref{tab:comparison_cga1a2} compares the coefficients and the corresponding absolute growth rates. At the linear critical condition $T_c=150.95$ (very close to the value 150.77 obtained from numerical simulations in Fig. \ref{fig.hystloop_varU}), a favourable agreement can be observed and the relative errors between the absolute growth rates are just 0.09\%. However, in terms of the multiple-scale expansion, with $T$ increased from $T_c$, the error increases rapidly, which arises mainly from the discrepancy between the linear growth rate $a_{1,r}(T-T_c)$ and $a_{1T,r}$; as for the amplitude expansion, the error increases much more slowly. Even at $T=300$, far away from the linear critical condition, the calculated absolute growth rate in the amplitude expansion method has only a relative error $0.52\%$ compared to the dispersion relation method. This illustrates that the coefficients evaluated based on the multiple-scale expansion indeed cannot be used for a reliable analysis when $T$ is far from its critical value, while the results of the amplitude expansion method remain quite accurate even far away from the linear critical condition. To the best of our knowledge, this appears to be a new result demonstrating the difference between the two expansion methods in the weakly nonlinear phase in terms of the absolute instability.

\begin{figure}
	\centering
	\includegraphics[width=1.0\textwidth]{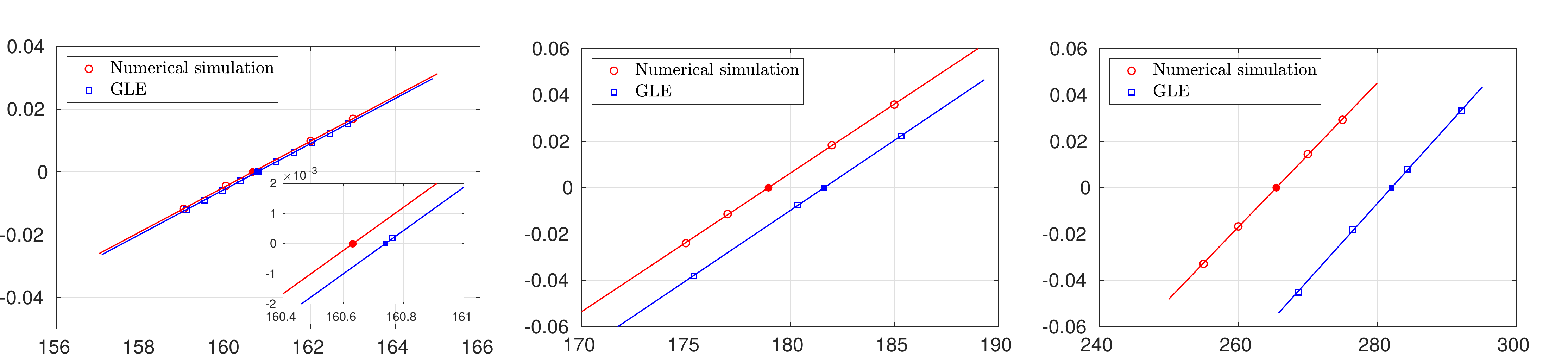}
	\put(-521,102){(a)}
	\put(-481,72){$\mathcal{U}=0.2$}
	\put(-515,53){$\sigma$}
	\put(-424,-7){$T$}
	\put(-351,102){(b)}
	\put(-311,72){$\mathcal{U}=0.5$}
	\put(-256,-7){$T$}
	\put(-182,102){(c)}
	\put(-144,72){$\mathcal{U}=1$}
	\put(-87,-7){$T$}
	\caption{Variation of the absolute growth rate $\sigma(v_g=0)$ with the electric Rayleigh number $T$ for EHD-Poiseuille flows at $C=10$, $M=10$ and $Fe=10^4$. The red circular symbols refer to current results computed by post-processing the snapshots from numerical simulations, blue square symbols are results predicted from the GLE, and solid lines denote the corresponding linearly fitted results with solid dots indicating $T_{ca}$. At $T<T_{ca}$, $\sigma<0$, the flow is convectively unstable; at $T>T_{ca}$, $\sigma>0$, the flow is absolutely unstable; at $T=T_{ca}$, $\sigma=0$, the flow is linearly neutral. (a) $\mathcal{U}=0.2$, $T_{ca}^{Simulation}=160.63$ and $T_{ca}^{GLE}=160.74$; (b) $\mathcal{U}=0.5$, $T_{ca}^{Simulation}=178.96$ and $T_{ca}^{GLE}=181.63$; (c) $\mathcal{U}=1$, $T_{ca}^{Simulation}=265.47$ and $T_{ca}^{GLE}=282.06$. }
	\label{fig.find_crit}
\end{figure}
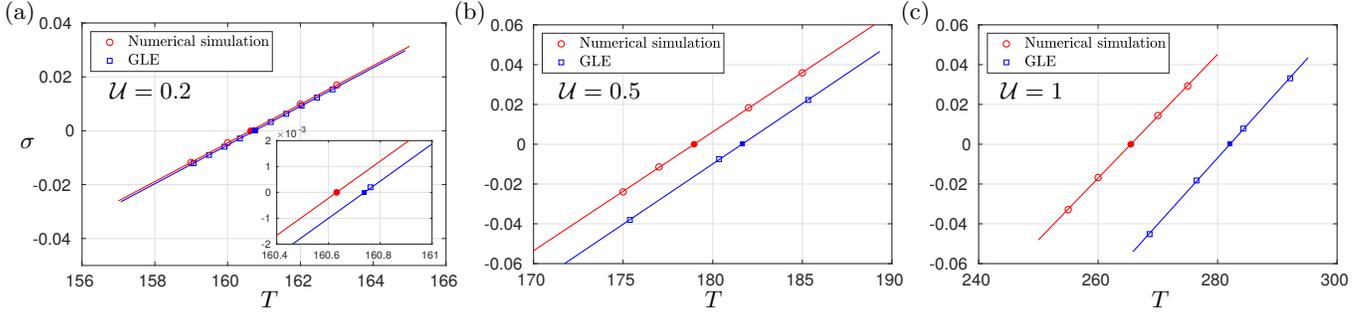

In order to gain a global picture of how the absolute growth rate varies with $T$, in Fig. \ref{fig.find_crit}, we plot the absolute growth rates $\sigma(v_g=0)$ as a function of $T$ for $\mathcal{U}=[0.2,0.5,1]$ (red circles). Plotted in the same figures are the absolute growth rate computed using the GLE (blue squares). The critical electric Rayleigh number $T_{ca}$ indicating the onset of linearly absolute instability is represented by solid symbols (when $T<T_{ca}$, convectively unstable and when $T>T_{ca}$, absolutely unstable), and they are computed by the linear least squares method. As the increase of through-flow $\mathcal{U}$, it can be clearly seen that the spatiotemporal instability criterion $T_{ca}$ also increases, which is consistent with the linear result in our previous work \citep{Li2019}.

At $\mathcal{U}=0.2$, the spatiotemporal instability criteria $T_{ca}$ predicted from current post-processing method of numerical simulations and the GLE model agree well with each other as observed from Fig. \ref{fig.find_crit}, and their difference is about $0.07\%$. While as the intensification of through-flow, the difference between above two results gradually increases to $1.49\%$ at $\mathcal{U}=0.5$ and $6.25\%$ at $\mathcal{U}=1$. This discrepancy could be attributed to the fact that the dispersion relation of equation (\ref{eq.epsilons}a) can not be approximated well enough by a quadratic function over the range of streamwise wavenumbers within which the linear growth rate is positive \citep{Suslov2004Stability}. For example, Fig. \ref{fig.validateT190}(b) illustrates the difference between the parabolic approximation (dashed curve) and the true dispersion relation (solid curve). Therefore, the validity of the linearised GLE is restricted, leading to a reduced predictive ability of the convective and absolute instabilities \citep{Suslov2004Stability}. In addition, as the GLE is normally derived around the linear instability criterion $T_c$ \citep{zhang2016weakly}, it is only valid in a small region near $T_c$, and their results may be misleading if $T$ is far from $T_c$. From Figs. \ref{fig.hystloop_varU} and \ref{fig.find_crit}, it can be seen that the deviation between linear instability criterion $T_c$ and spatiotemporal instability criterion $T_{ca}$ becomes larger as the increase of through-flow strength $\mathcal{U}$.

\begin{figure}
  \centering
  \includegraphics[width=0.9\textwidth]{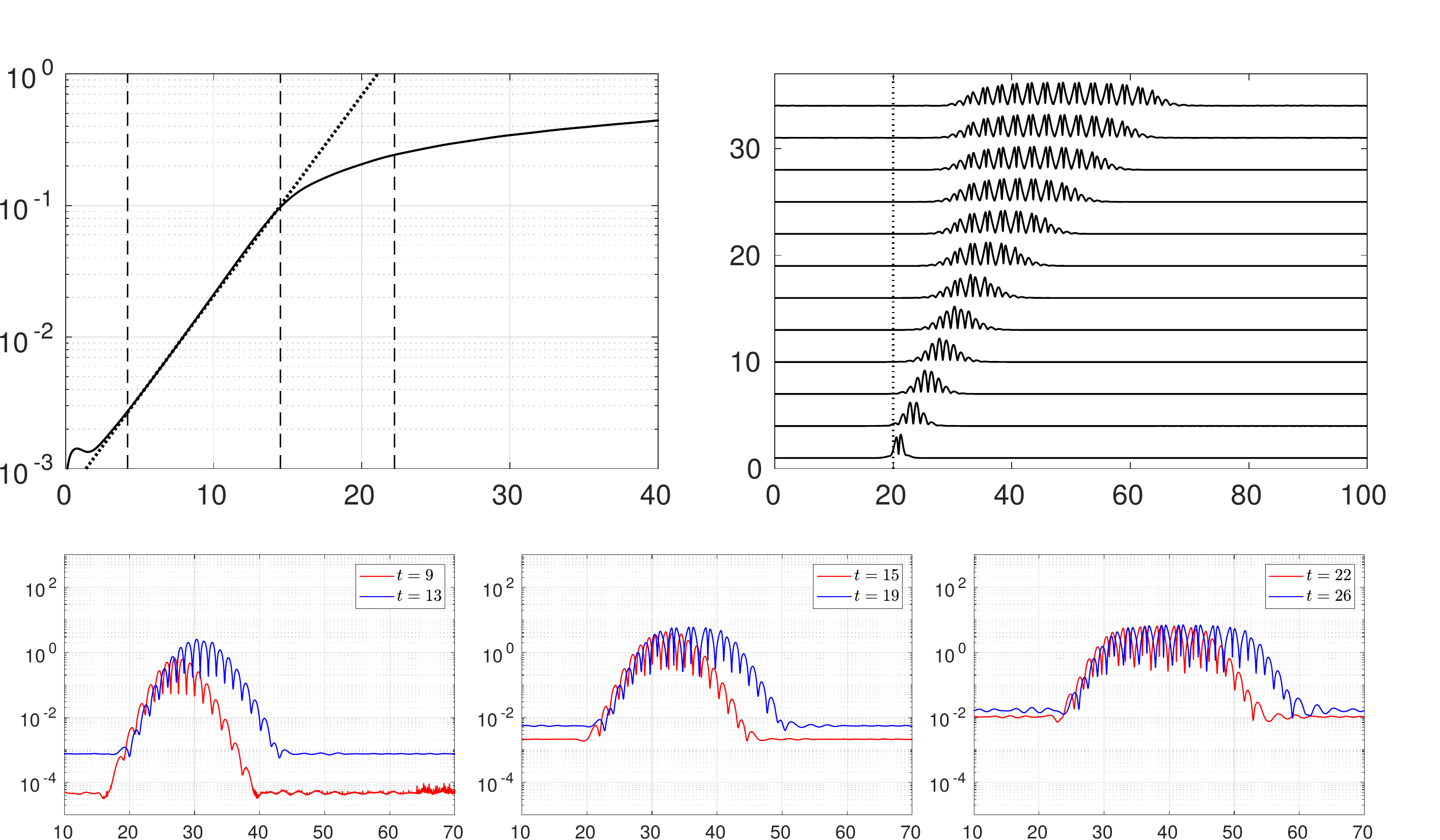}
  \put(-472,243){(a)}
  \put(-472,175){$\mathcal{A}$}
  \put(-345,99){$t$}
  \put(-433,187){\rotatebox{90}{transient}}
  \put(-405,187){\rotatebox{90}{exponential}}
  \put(-395,187){\rotatebox{90}{growth}}
  \put(-358,150){\rotatebox{90}{saturation}}
  \put(-300,150){\rotatebox{90}{wavepacket}}
  \put(-290,150){\rotatebox{90}{spreading}}
  \put(-242,243){(b)}
  \put(-238,175){$\eta$}
  \put(-118,99){$x$}
  \put(-50,123){$t=1$}
  \put(-50,154){$t=10$}
  \put(-50,185){$t=19$}
  \put(-50,215){$t=28$}
  \put(-457,86){(c)}
  \put(-460,47){$\eta$}
  \put(-379,-6){$x$}
  \put(-434,82){exponential growth}
  \put(-312,86){(d)}
  \put(-233,-6){$x$}
  \put(-287,82){saturation}
  \put(-166,86){(e)}
  \put(-88,-6){$x$}
  \put(-145,82){wavepacket spreading}
  \caption{Temporal evolution of the global energy amplitude $\mathcal{A}$ and the enstrophy $\eta$ describing the envelope of wavepackets for EHD-Poiseuille flows at $C=10$, $M=10$, $Fe=10^4$, $\mathcal{U}=1$ and $T=220$. (a) temporal evolutions of $\mathcal{A}$, it consists of four stages: transient; exponential growth of amplitude; saturation; wavepacket spreading; (b) waterfall of the wavepacket denoted by normalized enstrophy, the dotted line denotes the initial position $x_0=20$ of the impulse; (c) wavepacket within the regime of exponential growth; (d) wavepacket within the regime of saturation; (e) wavepacket within the regime of streamwise spreading.}
 \label{fig.ehd_TempEvol}
\end{figure}
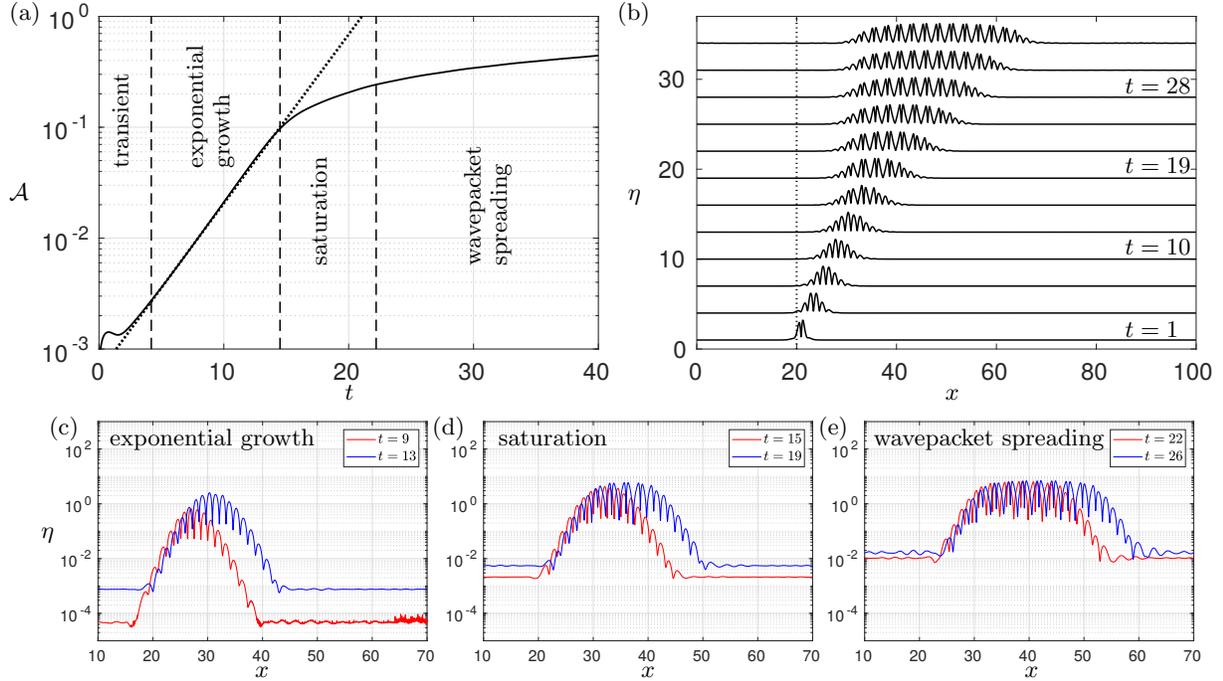

\subsubsection{Nonlinear impulse response}
When the initial local impulse becomes finite-amplitude and finite-extent, nonlinearity becomes important. The complete temporal evolution of the global energy amplitude $\mathcal{A}$ is shown in Fig. \ref{fig.ehd_TempEvol}(a). The envelopes of wavepackets within linear and nonlinear regimes are represented by the enstrophy $\eta$ defined in equation (\ref{eq.eta}), whose spatiotemporal evolution is described by the waterfall plot in panel b. The wavepacket at each instant is normalized by its maximum amplitude for a clear visualization. In panel a, after an initial transient period, the local impulse starts to grow exponentially as indicated by the linear dotted line. Within this linear range, the amplitude and extent of the impulse both grow with time as shown in panel b. After this linear regime, the wavepacket undergoes a saturation process, during which the growth rate of global amplitude $\mathcal{A}$ decreases. Such a phenomenon can also be clearly demonstrated in panel c, where the increment of impulse amplitude after every $\Delta t=4$ becomes smaller. Finally, in the regime of wavepacket spreading, the growth rate of $\mathcal{A}$ becomes even smaller because the impulse amplitude stops to grow and it only expands along the streamwise direction as shown in panel d. At $T=220<T_{ca}=265.47$, the EHD-Poiseuille flow is convectively unstable. Thus, the wavepacket overall moves to the downstream region as shown in these panels.

\begin{figure}
	\centering
	\includegraphics[width=0.7\textwidth]{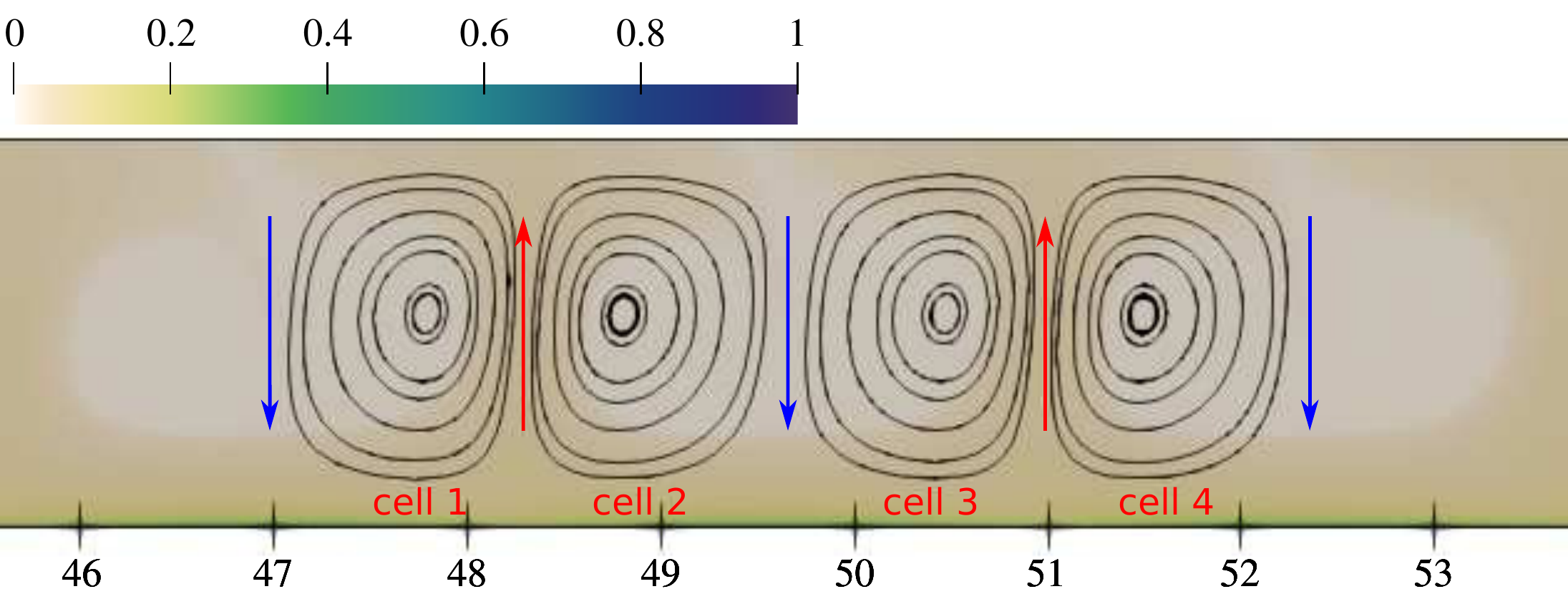}
	\put(-182,-7){$x$}
	\put(-367,59){$y$}
	\caption{Distribution of charge density and streamline pattern of perturbated velocity $\mathbf{u}$ for EHD-Poiseuille flows at $C=10$, $M=10$, $Fe=10^4$, $\mathcal{U}=1$ and $T=220$. Only part of the flow field is presented here for a clear visualisation. This snapshot corresponds to the wavepacket at $t=34$ in above Fig. \ref{fig.ehd_TempEvol}. Red arrows represent the upward motion of charges, and high-charge regions form around them. Blue arrows indicate downward moving charges with low density. }
	\label{fig.ehd_streamline_T220}
\end{figure}

\begin{figure}
	\centering
	\includegraphics[width=0.9\textwidth]{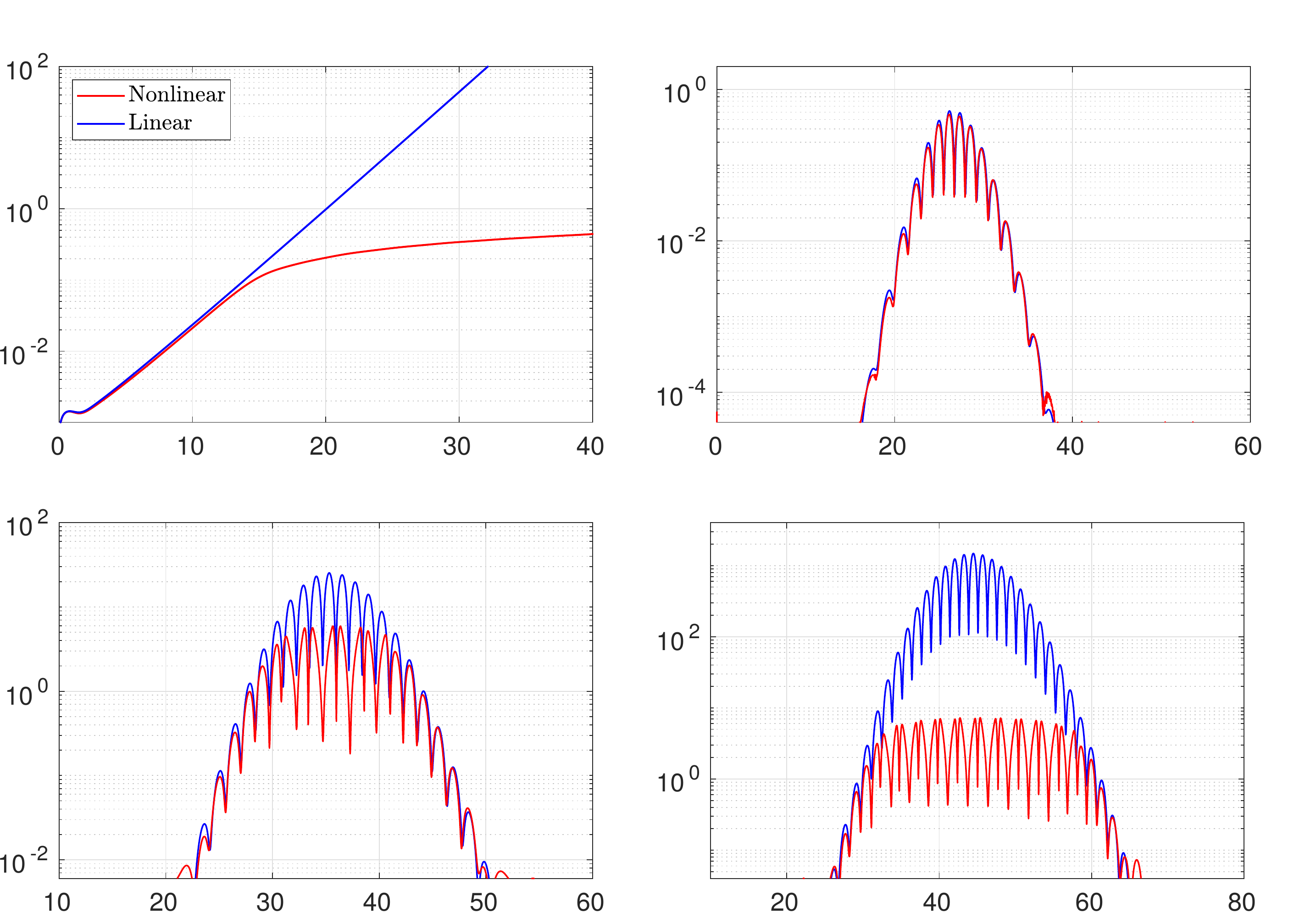}
	\put(-472,291){(a)}
	\put(-472,230){$\mathcal{A}$}
	\put(-347,150){$t$}
	\put(-242,291){(b)}
	\put(-238,230){$\eta$}
	\put(-118,150){$x$}
	\put(-198,280){$t=8$}
	\put(-472,130){(c)}
	\put(-467,65){$\eta$}
	\put(-346,-8){$x$}
	\put(-430,120){$t=19$}
	\put(-242,130){(d)}
	\put(-237,65){$\eta$}
	\put(-119,-8){$x$}
	\put(-200,120){$t=30$}
	\caption{Comparison of the global energy amplitude $\mathcal{A}$ and the enstrophy $\eta$ between linear (blue solid lines) and nonlinear (red solid lines) cases for EHD-Poiseuille flows at $C=10$, $M=10$, $Fe=10^4$, $\mathcal{U}=1$ and $T=220$. The EHD-Poiseuille flow is convectively unstable. (a) temporal evolutions of $\mathcal{A}$; (b) $t=8$ within regime of exponential growth; (c) $t=19$ within regime of saturation; (d) $t=30$ within regime of wavepacket spreading.}
	\label{fig.ehd_nlinRes_T220}
\end{figure} 

Within the regime of wavepacket spreading in panels b and e of Fig. \ref{fig.ehd_TempEvol}, it can be seen that those local maxima peaks are not equally spaced. At $T>T_c$, the electroconvective flow takes place around the upward moving charges (they form the high-charge region) or downward moving charges (they form the low-charge region) as shown in Fig. \ref{fig.ehd_streamline_T220}. As the high-charge region is much thinner than the low-charge one, the streamwise distance between centers of cell 1 and 2 ($\Delta x_{12}=1$) is smaller than that between cells 2 and 3 ($\Delta x_{23}=1.7$). Thus, the enstrophy $\eta$ after saturation regime does not present the uniformly distributed maxima peaks. From a point of view of flow physics, it is because of the charge-void regions (see the almost white regions in Fig. \ref{fig.ehd_streamline_T220}) and the flow field that causes the non-uniform peaks in EHD-Poiseuille flows. This is different from the case in RBC subjected to a Poiseuille flow, see the results in the Supplemental Material. 

Next, we will compare the impulse response in linear and nonlinear regimes. When starting from the same initial condition consisting of one local impulse, the time series of global energy amplitude $\mathcal{A}$ obtained from nonlinear equations (\ref{eq.nlinehd}) and linear equations (\ref{eq.linehd}) at $T=220$ are compared in Fig. \ref{fig.ehd_nlinRes_T220}. Within the linear stage, the global amplitudes agree well between the linear and nonlinear simulations. With the increase of the perturbation amplitude, the deviation of the nonlinear simulation from the linear one becomes larger. The envelopes of the wavepackets in linear and nonlinear regimes at different instants are compared using enstrophy $\eta$ in panels b, c and d. It makes sense that the wavepackets coincide with each other in the linear stage (at $t=8$), as shown in panel a. At the subsequent instants, the mismatch between linear and nonlinear results becomes more visible, but interestingly, the leading and trailing edges in linear (moves at velocity $v_{\pm}$) and nonlinear (moves at velocity $v_{\pm}^{NL}$) simulations are still roughly the same after the period of linear growth, which means that the front separating the saturated wavepacket from the basic flow moves at the same speed in the linear and nonlinear regimes (i.e. $v_-=v_-^{NL}$ and $v_+=v_+^{NL}$). At $T=300$, the EHD-Poiseuille flow becomes absolutely unstable based on the conclusion from Fig. \ref{fig.find_crit}, and it is also found that the wavepackets obtained from linear and nonlinear equations at the same instant moves at the same speed (results now shown due to space limit). This conclusion is also valid for $\mathcal{U}=0.2$ and $\mathcal{U}=0.5$ (not shown due to space limit). These results demonstrate that the linearly convective/absolute instability criteria can be used to predict the transition from nonlinearly convective to absolute instability, at least for the EHD-Poiseuille flow that we have studied here. Due to the similar boundary conditions and flow patterns with EHD-Poiseuille flows, the results of nonlinear impulse response in RBC-Poiseuille flows are also studied and provided in the Supplemental Material. In the latter flow, the nonlinearity also does not influence the edge velocities of wavepackets.

\begin{figure}
  \centering
  \includegraphics[width=0.9\textwidth]{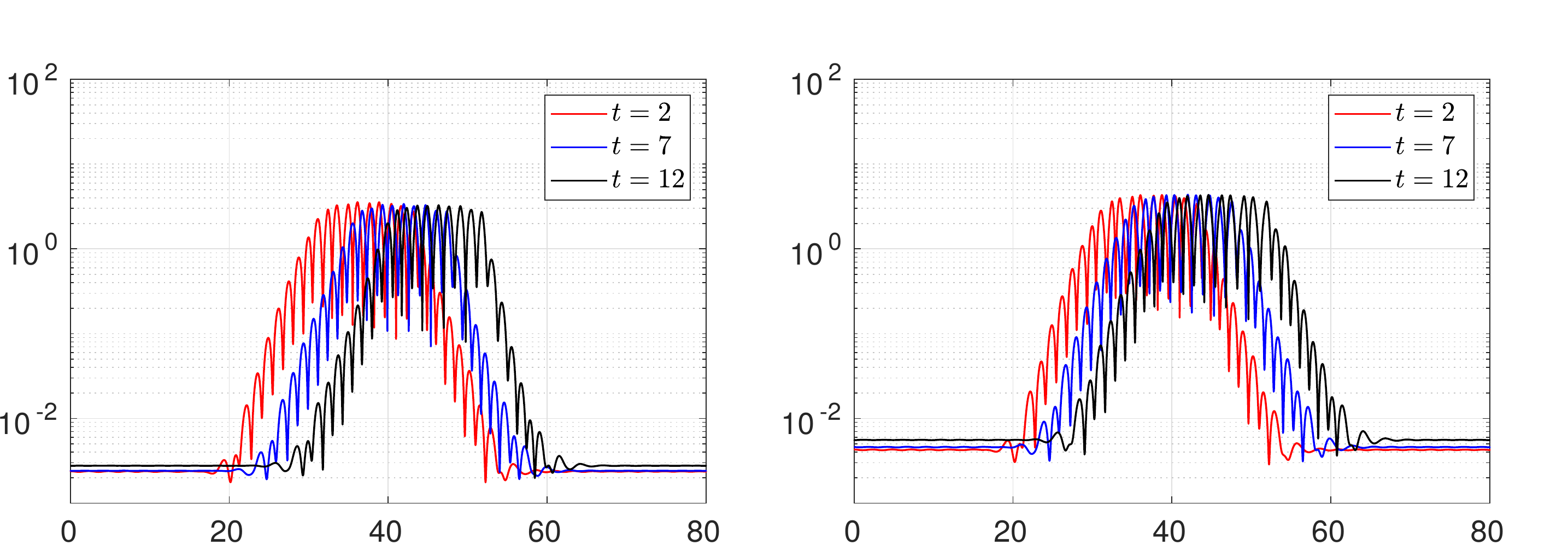}
  \put(-472,130){(a)}
  \put(-467,65){$\eta$}
  \put(-346,-8){$x$}
  \put(-430,120){$T=125$}
  \put(-242,130){(b)}
  \put(-237,65){$\eta$}
  \put(-119,-8){$x$}
  \put(-200,120){$T=145$}
  \caption{Temporal evolution of the enstrophy $\eta$ within the finite-amplitude range $T\in[T_f,T_c]$ for EHD-Poiseuille flows at $C=10$, $M=10$, $Fe=10^4$ and $\mathcal{U}=1$. (a) $T=125$; (b) $T=145$.}
 \label{fig.ehd_nlinRes_finite}
\end{figure}

\begin{figure}
  \centering
  \includegraphics[width=0.7\textwidth]{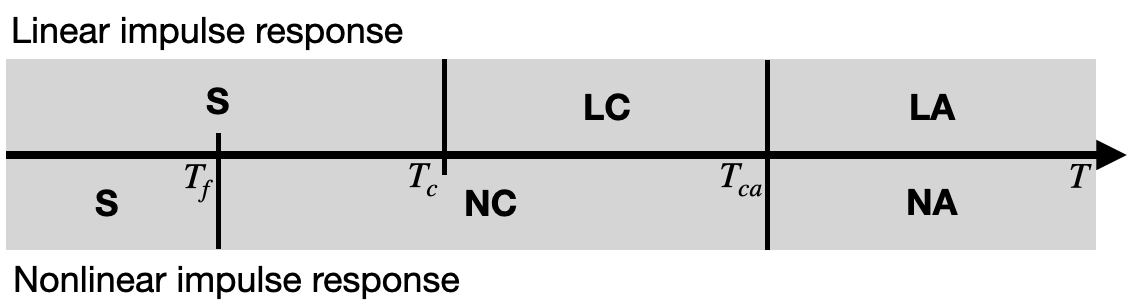}
  \caption{Linear and nonlinear impulse responses of the EHD-Poiseuille flow at $C=10$, $M=10$ and $Fe=10^4$. S: stable; LC: linearly convective; LA: linearly absolute; NC: nonlinearly convective; NA: nonlinearly absolute. $T_f$: finite-amplitude instability criterion; $T_c$: linear instability criterion; $T_{ca}$: instability criterion between LC and LA. Coexistence of $T_c$ and $T_f$ indicates the subcritical nature of bifurcations. }
 \label{fig.summary_aici_ehd}
\end{figure}

It is known that the bifurcation of EHD-Poiseuille flows is subcritical, and a hysteresis loop exists in the bifurcation diagram (see Fig. \ref{fig.hystloop_varU}). Thus, within the parameter range $T\in[T_f,T_c]$, an initially nonlinear finite-amplitude and finite-extent impulse should evolve without decaying to the steady state. Different from the linear impulse explicitly given by equation (\ref{eq.impulse}), the ultimately saturated solution of EHD-Poiseuille flows at $T=160$ (larger than $T_c$) can be directly adopted as the initially nonlinear impulse (i.e. initial condition) of EHD-Poiseuille flows within the finite-amplitude regime. In order to understand the instability nature of these finite-amplitude impulses, the spatiotemporal evolutions of nonlinear wavepackets at $T=125$ and $T=145$ are shown in Fig. \ref{fig.ehd_nlinRes_finite}. At the same instant, the wavepacket envelope of $T=145$ is wider than that of $T=125$ due to the stronger electroconvection in former one. As the increase of time, the variation of impulse amplitude is very small, while the overall wavepacket moves to the downstream region which clearly indicates the nonlinearly convective instability of EHD-Poiseuille flows within the finite-amplitude range $T\in[T_f,T_c]$. In the end, the above linear and nonlinear results are summarised in Fig. \ref{fig.summary_aici_ehd}. The difference between linear and nonlinear impulse responses stems from the characteristic of subcritical bifurcation in EHD-Poiseuille flows. 

\begin{figure}
	\centering
	\includegraphics[width=0.5\textwidth]{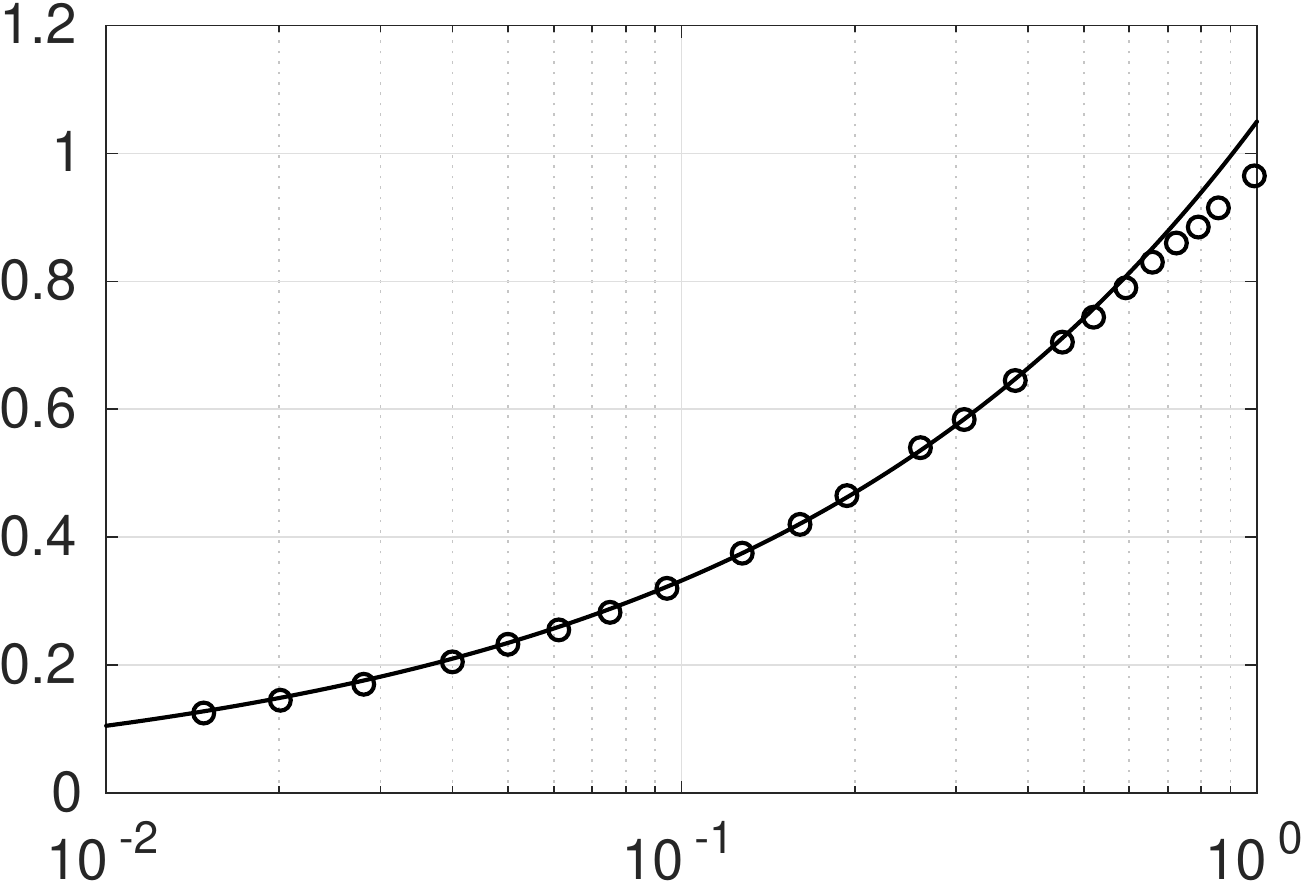}
	\put(-123,-10){$\epsilon'$}
	\put(-270,90){$c_f$}
	\caption{Variation of the front-propagation velocity $c_f$ with the parameter $\epsilon'=(T-T_c)/T_c$ for EHD-Poiseuille flows at $C=10$, $M=10$, $Fe=10^4$ and $\mathcal{U}=1$. The circular symbols refer to the results of numerical simulations, and solid line comes from the theoretical prediction $c_f \approx \sqrt{\epsilon'}$.}
	\label{fig.ehd_front}
\end{figure}

In Fig. \ref{fig.ehd_TempEvol}, after the stage of linear growth, the initially localised impulse finally evolves to the saturated state with finite amplitude and extent, and the linear edges can transform into nonlinear fronts \citep{van1992fronts}, which indicates the formation of electroconvection in downstream regions of the background through-flow. Such front separating the stable and unstable regimes can propagate with a well-defined velocity, namely the front-propagation velocity $c_f$ \citep{fineberg1987vortex,van2003front}. This velocity quantifies flow state of the wavepacket, and is related to the linear edge velocities $v_{\pm}$ shown in Figs. \ref{fig.tamp_vg}(c-d). Their difference is that the former is computed in the reference frame of the background through-flow, and the latter are obtained in the laboratory frame. In the current work, the background through-flow (Poiseuille flow) does not change with time, and edges of wavepackets from linear to nonlinear regimes always coincide with each other (see Fig. \ref{fig.ehd_nlinRes_T220}). Thus, the front-propagation velocity can be computed using the leading and trailing velocities of linear wavepackets, namely $c_f=(v_{+}-v_{-})/2$. The variation of $c_f$ with the control parameter $\epsilon'=(T-T_c)/T_c$ (which is similar to $\epsilon$ in the weakly nonlinear stability analysis in Eqs. \ref{eq.expansion-tzgammaT}) is shown in Fig. \ref{fig.ehd_front}, and $\epsilon'$ also describes the deviation of $T$ from its linear instability criterion ($T_c=150.77$ in Fig. \ref{fig.ehd_front}). Such relation can also be predicted from an Ginzburg-Landau equation, which indicates that the $c_f \propto \xi \sqrt{\epsilon'}$ \citep{dee1983propagating} (with $\xi$ being a certain pre-factor). In different flows, the value of $\xi$ may vary  \citep{ahlers1983vortex}. For the EHD-Poiseuille flow, near the linear instability criterion in Fig. \ref{fig.ehd_front}, it can be seen that the front-propagation velocity approximately follow the relation of $c_f \approx \sqrt{\epsilon'}$, while the deviation increases at high $\epsilon'$.

\vspace{-0.5cm}
\section{Conclusion and Future Work} \label{sec:conclusion}

In this paper, we investigated the spatiotemporal instabilities and weakly nonlinear properties of the EHD-Poiseuille flows by a post-processing method of highly accurate numerical simulations. We aim to characterise the linear and nonlinear AI/CI in EHD-Poiseuille flows, as well as to test the potential of GLE in predicting the flow nature around its first bifurcation and the transition from convective to absolute instability. As the EHD flow is subjected to a through-flow, the effect of this through-flow strength is also studied. Our new findings are summarised below.

In the weakly nonlinear analysis of EHD-Poiseuille flows, we use three methods to compute the Landau coefficient $c_3$ which indicates the nature of the first bifurcation. The first two methods are based on the amplitude equation, derived from two different expansion methods, namely the multiple-scale method and amplitude expansion method. The third method computes the Landau coefficient from the results of numerical simulations. Near the linear instability threshold, the Landau coefficients computed from these three methods are all positive, confirming the subcritical bifurcation of EHD-Poiseuille flows. As the increase of electric Rayleigh number $T$, the values of the Landau coefficient obtained in the amplitude expansion method agree well with those from the post-processing method, but the deviation of $c_3$ becomes larger between the multiple-scale method and the post-processing method. This clearly indicates the limited convergence range of the multiple-scale expansion method and the applicability of the amplitude expansion method, even when $T$ is away from the linear instability threshold. Using the highly accurate simulations, the subcritical bifurcation diagrams of EHD-Poiseuille flows with different through-flow strengths $\mathcal{U}$ were also computed with a diagnosis of the electric Nusselt number $Ne$. As the increase of $\mathcal{U}$, the linear instability criterion $T_c$ decreases while the finite-amplitude criterion $T_f$ increases, which indicates a narrower hysteresis loop.

By studying the response of a localised impulse in the flow, we extracted information on the spatiotemporal instabilities of EHD-Poiseuille flows from the snapshots of numerical simulations. The GLE has also been derived to predict the transition from CI to AI in order to test its predicability. When the through-flow is weak, as the spatiotemporal transition takes place near the linear critical condition $T_c$, the spatiotemporal instability criterion $T_{ca}$ predicted by the GLE agrees well with that from the post-processing method of numerical simulations. While as the increase of $\mathcal{U}$, $T_{ca}$ becomes larger than $T_c$, and the applicability of GLE becomes increasingly questionable. On the other hand, when $\mathcal{U}$ is fixed, by changing $T$, we found that the GLE coefficients calculated using the amplitude expansion method can yield absolute growth rates very close to those obtained from the dispersion relation, a conclusion which does not hold for the multiple-scale expansion method, whose predicted absolute growth rates are only close to those computed by the dispersion relation near the linear critical conditions. This result demonstrates the difference between the two expansion methods in the weakly nonlinear phase in terms of the absolute growth rate.

Because of the nonlinearity in the flow, the localised impulse will, after the linear growth phase, eventually evolve into a saturated wavepacket. It is found that edges of the impulse and saturated wavepackets always coincide with each other, which means that the nonlinearity does not influence the propagation velocity of the wavepacket except for its amplitude (i.e., the saturation). The same conclusion is also found for the RBC-Poiseuille flows in the appendix. Within the finite-amplitude range $T\in[T_f,T_c]$, the spatiotemporal property of EHD-Poiseuille flows is proved to be nonlinearly convective at least for the parameters we have investigated in this work. In the nonlinear evolution process of the wavepacket, it is also found that the front-propagation velocity increases with the control parameter $\epsilon'=(T-T_c)/T_c$ according to a power law (i.e. $c_f \propto \sqrt{\epsilon'}$), which is consistent with the result predicted by amplitude equations.

The current work studies the nonlinear spatiotemporal development of the disturbance in EHD-Poiseuille flows on, e.g., its transition from nonlinear CI to nonlinear AI and the power law for the front-propagation velocity. These results can potentially provide some useful implications for studying the fluid dynamics in industrial processes like ESP and EV damper devices. To this end, the same flow configuration as that, for example, of ESP (wires between two plates) should be used. 
Besides, here we mainly study the effect of electric Rayleigh number on the nonlinear AI/CI, future works can consider a more complete parameter study (including the injection strength $C$ and mobility ratio $M$) and explore a larger parameter space in this flow. Finally, since now we have demonstrated that the EHD flow can be influenced by the Poiseuille flow, it may also be interesting to study how the EHD rolls can change the Poiseuille flow, for example, delaying the turbulence transition in the Poiseuille flow \cite{soldati1998turbulence} from the perspective of flow stability.

\begin{acknowledgments}
We acknowledge the financial support of a Start-up grant from the Ministry of Education, Singapore to M.Z. (with the WBS No. R-265-000-689-114) and doctoral research scholarships from the National University of Singapore to Z.F. and D.W. D.W. is also partially supported by the China Scholarship Council. The computational resources of the National Supercomputing Centre, Singapore (https://www.nscc.sg) are acknowledged.
\end{acknowledgments}

\appendix
\section{Operators in the weakly nonlinear analysis} \label{sec:append_operators}
In this appendix, we provide the explicit expressions of the various operators appearing in the weakly nonlinear analysis, including those in section \ref{sec:multiple-scale expansion} for multiple-scale expansion and those in section \ref{sec:amplitude_expansion} for amplitude expansion. All the below expressions are completely consistent with those in our previous study \citep{zhang2016weakly} (see the appendices therein); the only difference is that in that paper the operators are given in physical space, while in the present work we express them directly in the spectral space.

\subsection{Weight matrices}
The weight matrix $\boldsymbol{M}$ in physical space introduced in equation (\ref{eq.compact-form}) is 
$\boldsymbol{M}= \begin{bmatrix}
\nabla^2 & 0 \\ 0 & \nabla^2
\end{bmatrix}$. To facilitate the expression of various operators related to it, we introduce in spectral space the operator 
$\tilde{\boldsymbol{M}}^{(n)}= \begin{bmatrix}
\nabla_n^2 & 0 \\ 0 & \nabla_n^2
\end{bmatrix}$
where $ \nabla_n^2=-n^2 \alpha^2 + ({\partial^2}/{\partial y^2}) $ and $ \nabla_n^4= \nabla_n^2 \nabla_n^2 $ for later use.
Then for the multiple-scale expansion in section \ref{sec:multiple-scale expansion}, the $\boldsymbol{M}$--related operators appearing in equation (\ref{eq.cg}) and (\ref{eq.Landau-coefficients}) can be expressed as $\tilde{\boldsymbol{M}}_0=\tilde{\boldsymbol{M}}^{(1)}$, 
$\tilde{\boldsymbol{M}}_1^{\circ}= \begin{bmatrix}
2i \alpha & 0 \\ 0 & 2i \alpha
\end{bmatrix}$ and
$\tilde{\boldsymbol{M}}_2^{\circ}= \begin{bmatrix}
1 & 0 \\ 0 & 1
\end{bmatrix}$.
For the amplitude expansion in section \ref{sec:amplitude_expansion}, the  $\boldsymbol{M}$--related operators appearing in equation (\ref{eq.operators}) are
$\tilde{\boldsymbol{M}}_{0\alpha}=\tilde{\boldsymbol{M}}^{(0)}$, $\tilde{\boldsymbol{M}}_{1\alpha}=\tilde{\boldsymbol{M}}^{(1)}$ and $\tilde{\boldsymbol{M}}_{2\alpha}=\tilde{\boldsymbol{M}}^{(2)}$; the operators in the expressions of $\tilde{\boldsymbol{n}}_{21}$ and $\tilde{\boldsymbol{n}}_{31z}$ in equation (\ref{eq.amplitude_equations}) are $\tilde{\boldsymbol{M}}_{1\alpha}^{\circ}=\tilde{\boldsymbol{M}}_1^{\circ}$ and $\tilde{\boldsymbol{M}}_{2\alpha}^{\circ}=\tilde{\boldsymbol{M}}_2^{\circ}$.

\subsection{Linear operators}
The linear operator $\boldsymbol{L}$ in equation (\ref{eq.compact-form}) is in the physical space and it is
\begin{equation}
\boldsymbol{L}= \begin{bmatrix}
-\bar{U} \frac{\partial  \nabla^2}{\partial x} + \bar{U}'' \frac{\partial }{\partial x} + \frac{M^2}{T} \nabla^4   & - M^2 \left(\bar{\phi}' \frac{\partial \nabla^2 }{\partial x} - \bar{\phi}''' \frac{\partial }{\partial x} \right) \\\bar{\phi}''' \frac{\partial }{\partial x}   & \bar{\phi}''' \frac{\partial }{\partial y} + \left( \bar{\phi}' \frac{\partial}{\partial y}-\bar{U}\frac{\partial}{\partial x} \right)\nabla^2  + 2 \bar{\phi}'' \nabla^2  + \frac{1}{Fe} \nabla^4 
\end{bmatrix}.
\end{equation}
Similar to $\tilde{\boldsymbol{M}}^{(n)}$, we introduce $\tilde{\boldsymbol{L}}^{(n)}$ for convenience
\begin{equation}
\tilde{\boldsymbol{L}}^{(n)}= \begin{bmatrix}
-\bar{U} in\alpha{\nabla_n^2} + \bar{U}'' in\alpha + \frac{M^2}{T} \nabla_n^4   & - M^2 \left(\bar{\phi}' in\alpha { \nabla_n^2 } - \bar{\phi}''' in\alpha \right) \\\bar{\phi}''' in\alpha   & \bar{\phi}''' \frac{\partial }{\partial y} + \left( \bar{\phi}' \frac{\partial}{\partial y}-\bar{U}in\alpha \right)\nabla_n^2  + 2 \bar{\phi}'' \nabla_n^2  + \frac{1}{Fe} \nabla_n^4 
\end{bmatrix}.
\end{equation}
Then the $\boldsymbol{L}$--related operators in equations (\ref{eq.cg}) and (\ref{eq.Landau-coefficients}) for the multiple-scale expansion read
$\tilde{\boldsymbol{L}}_0=\tilde{\boldsymbol{L}}^{(1)}$, and
\begin{equation}
\tilde{\boldsymbol{L}}_1^{\circ}= \begin{bmatrix}
-\bar{U} (\nabla_1^2 -2\alpha^2) + \bar{U}'' + \frac{4 i \alpha M^2}{T_c}  \nabla_1^2   & - M^2 \left(\bar{\phi}' \nabla_1^2 -2 \alpha^2 \bar{\phi}'   - \bar{\phi}'''  \right) \\\bar{\phi}'''    &  2 i \alpha\left( \bar{\phi}' \frac{\partial}{\partial y} + 2 \bar{\phi}''   \right)   - \bar{U}(\nabla_1^2 -2 \alpha^2)  + \frac{4 i \alpha}{Fe}  \nabla_1^2 
\end{bmatrix},
\end{equation}
\begin{equation}
\tilde{\boldsymbol{L}}_2^{\circ}= \begin{bmatrix}
- 3 i \alpha\bar{U} +  \frac{M^2}{T_c} (2 \nabla_1^2 -4\alpha^2)   & -3 i \alpha M^2 \bar{\phi}'    \\0   &  \left( \bar{\phi}' \frac{\partial}{\partial y} + 2 \bar{\phi}''   \right)   -3 i \alpha \bar{U}   + \frac{1}{Fe} ( 2 \nabla_1^2-4 \alpha^2)
\end{bmatrix},  \,
\tilde{\boldsymbol{L}}_{2MT}= \begin{bmatrix}
- \frac{M^2}{T_c^2} \nabla_1^4  & 0 \\0    &  0 
\end{bmatrix}.
\end{equation}
Those appearing in equation (\ref{eq.operators}) for the amplitude expansion are
$\tilde{\boldsymbol{L}}_{0\alpha}=\tilde{\boldsymbol{L}}^{(0)}$, $\tilde{\boldsymbol{L}}_{1\alpha}=\tilde{\boldsymbol{L}}^{(1)}$ and $\tilde{\boldsymbol{L}}_{2\alpha}=\tilde{\boldsymbol{L}}^{(2)}$. The operators in the expressions of $\tilde{\boldsymbol{n}}_{21}$ and $\tilde{\boldsymbol{n}}_{31z}$ in equation (\ref{eq.amplitude_equations}) are $\tilde{\boldsymbol{L}}_{1\alpha}^{\circ}=\tilde{\boldsymbol{L}}_1^{\circ}$ and $\tilde{\boldsymbol{L}}_{2\alpha}^{\circ}=\tilde{\boldsymbol{L}}_2^{\circ}$.

\subsection{Nonlinear operators}
The nonlinear operator in physical space as introduced in equation (\ref{eq.compact-form}) is
${\boldsymbol{N}}=( n_{\psi},n_{\varphi})^T$ with
\begin{subequations}
	\begin{align}
	& n_{\psi} =  \left(  \frac{\partial \psi}{\partial x} \frac{\partial \nabla^2 \psi}{\partial y} -\frac{\partial \psi}{\partial y} \frac{\partial \nabla^2 \psi}{\partial x} \right) + M^2 \left(  \frac{\partial \varphi}{\partial x} \frac{\partial \nabla^2 \varphi}{\partial y} -\frac{\partial \varphi}{\partial y} \frac{\partial \nabla^2 \varphi}{\partial x} \right), \\
	&n_{\varphi} = \left( \frac{\partial \varphi}{\partial x} - \frac{\partial \psi}{\partial y}  \right) \frac{\partial \nabla^2 \varphi}{\partial x} + \left( \frac{\partial \varphi}{\partial y} + \frac{\partial \psi}{\partial x}  \right) \frac{\partial \nabla^2 \varphi}{\partial y} + \nabla^2 \varphi \nabla^2 \varphi.
	\end{align}
\end{subequations}
The nonlinear operator $\tilde{\boldsymbol{N}}_3^{\circ}$ appearing in equation (\ref{eq.Landau-coefficients}) for the multiple-scale expansion is
\begin{equation}
\tilde{\boldsymbol{N}}_3^{\circ} = \widetilde{\boldsymbol{N}}_f (\tilde{\boldsymbol{\gamma}}_{1},\tilde{\boldsymbol{\gamma}}_{20},1,0) + \widetilde{\boldsymbol{N}}_f (\tilde{\boldsymbol{\gamma}}_{20},\tilde{\boldsymbol{\gamma}}_{1},0,1) + \widetilde{\boldsymbol{N}}_f (\tilde{\boldsymbol{\gamma}}_{1}^*,\tilde{\boldsymbol{\gamma}}_{22},-1,2) + \widetilde{\boldsymbol{N}}_f (\tilde{\boldsymbol{\gamma}}_{22},\tilde{\boldsymbol{\gamma}}_{1}^*,2,-1),
\end{equation}
where the superscript ${}^*$ denotes the complex conjugate of its argument and the function $\widetilde{\boldsymbol{N}}_f$ is defined as $\widetilde{\boldsymbol{N}}_f (\tilde{\boldsymbol{f}}_1,\tilde{\boldsymbol{f}}_2,q_1,q_2)=(\tilde n_{\psi},\tilde n_{\varphi})^T$ with
\begin{subequations}
	\begin{align}
	&\tilde n_{\psi}(\tilde{\boldsymbol{f}}_1,\tilde{\boldsymbol{f}}_2,q_1, q_2) = \left(iq_1 \alpha \tilde{f}_{1\psi} \frac{\partial \nabla_{q_2}^2 \tilde{f}_{2\psi}}{\partial y} - \frac{\partial \tilde{f}_{1\psi}}{\partial y} i q_2 \alpha \nabla_{q_2}^2 \tilde{f}_{2\psi}  \right) + M^2 \left( iq_1 \alpha \tilde{f}_{1\varphi} \frac{\partial \nabla_{q_2}^2 \tilde{f}_{2\varphi}}{\partial y} -\frac{\partial \tilde{f}_{1\varphi}}{\partial y} i q_2 \alpha \nabla_{q_2}^2 \tilde{f}_{2\varphi} \right), \\
	&\tilde n_{\varphi}(\tilde{\boldsymbol{f}}_1,\tilde{\boldsymbol{f}}_2,q_1,q_2) = \left( iq_1\alpha\tilde{f}_{1\varphi} - \frac{\partial \tilde{f}_{1\psi}}{\partial y}  \right) i q_2 \alpha \nabla_{q_2}^2 \tilde{f}_{2\varphi} + \left( \frac{\partial \tilde{f}_{1\varphi}}{\partial y} + iq_1\alpha\tilde{f}_{1\psi}  \right) \frac{\partial \nabla_{q_2}^2 \tilde{f}_{2\varphi}}{\partial y} + \nabla_{q_1}^2 \tilde{f}_{1\varphi} \nabla_{q_2}^2 \tilde{f}_{2\varphi}.
	\end{align}
\end{subequations}
Here, the second subscripts in $\tilde f_1$ and $\tilde f_2$ label the corresponding components in the column vectors $\tilde{\boldsymbol{f}}_1=(\tilde{f}_{1\psi},\tilde{f}_{1\varphi})^T$ and $\tilde{\boldsymbol{f}}_2=(\tilde{f}_{2\psi},\tilde{f}_{2\varphi})^T$.
That in equation (\ref{eq.gamma31_K}) for the amplitude expansion is in the same form as $\tilde{\boldsymbol{N}}_3^{\circ}$ (but with different notations to differentiate the two expansion methods)
\begin{equation}
\tilde{\boldsymbol{N}}_{31} = \widetilde{\boldsymbol{N}}_f (\tilde{\boldsymbol{\gamma}}_{11},\tilde{\boldsymbol{\gamma}}_{20},1,0) + \widetilde{\boldsymbol{N}}_f (\tilde{\boldsymbol{\gamma}}_{20},\tilde{\boldsymbol{\gamma}}_{11},0,1) + \widetilde{\boldsymbol{N}}_f (\tilde{\boldsymbol{\gamma}}_{11}^*,\tilde{\boldsymbol{\gamma}}_{22},-1,2) + \widetilde{\boldsymbol{N}}_f (\tilde{\boldsymbol{\gamma}}_{22},\tilde{\boldsymbol{\gamma}}_{11}^*,2,-1).
\end{equation}
We also provide the corresponding expressions of the other two nonlinear terms used in the solution process: $\tilde{\boldsymbol{N}}_{20} = \widetilde{\boldsymbol{N}}_f (\tilde{\boldsymbol{\gamma}}_{11},\tilde{\boldsymbol{\gamma}}_{11}^*,1,-1) + \widetilde{\boldsymbol{N}}_f (\tilde{\boldsymbol{\gamma}}_{11}^*,\tilde{\boldsymbol{\gamma}}_{11},-1,1)$ and $\tilde{\boldsymbol{N}}_{22} = \widetilde{\boldsymbol{N}}_f (\tilde{\boldsymbol{\gamma}}_{11},\tilde{\boldsymbol{\gamma}}_{11},1,1)$.

\subsection{Adjoint operators}
The weight matrix in the adjoint equation is self-adjoint
$\tilde{\boldsymbol{M}}_0^{\dagger}= \tilde{\boldsymbol{M}}_0$. The adjoint linear operator is
\begin{equation}
\tilde{\boldsymbol{L}}_{0}^{\dagger}= \begin{bmatrix}
i \alpha \bar{U} \nabla_1^2+2i\alpha \bar{U}' \frac{\partial}{\partial y} + \frac{M^2}{T_c} \nabla_1^4 & -\kappa^2 \bar{\phi}''' i \alpha \\ \frac{M^2}{\kappa^2} [i\alpha (\bar{\phi}''' + 2 \bar{\phi}'' \frac{\partial}{\partial y} + \bar{\phi}'\frac{\partial^2}{\partial y^2}  ) - i \alpha^3 \bar{\phi}' - i \alpha \bar{\phi}''']    &  S 
\end{bmatrix},
\end{equation}
where the coefficient $\kappa$ comes from the inner product in equation (\ref{eq.inner-product-definition}) and the notation $S$ is short for
\begin{equation}
S = i \alpha \bar{U} \nabla_1^2 + i \alpha (\bar{U}'' + 2 \bar{U}' \frac{\partial}{\partial y}) - (\bar{\phi}'''' + \bar{\phi}''' \frac{\partial}{\partial y}) - (\bar{\phi}'\nabla_1^2 \frac{\partial }{\partial y} + 2\bar{\phi}'' \frac{\partial^2}{\partial y^2} +  \bar{\phi}''' \frac{\partial}{\partial y})  + (\bar{\phi}'' \nabla_1^2 + 2 \bar{\phi}''' \frac{\partial}{\partial y} + \bar{\phi}'''') + \frac{1}{Fe} \nabla_1^4.
\end{equation}
The adjoint eigenvector is $\tilde{\boldsymbol{\gamma}}_1^{\dagger}=(\tilde{\psi}_1^{\dagger}, \tilde{\varphi}_1^{\dagger})^T$. The boundary conditions for $\tilde{\psi}_1^{\dagger}$ and $\tilde{\varphi}_1^{\dagger}$ can be derived (from the integration by part process) as
\begin{subequations}
	\begin{equation}
		\tilde{\psi}_1^{\dagger} (\pm 1)=\frac{\partial \tilde{\psi}_1^{\dagger}}{\partial y} (\pm 1)=0,
	\end{equation}
	\begin{equation}
		\tilde{\varphi}_1^{\dagger}(\pm 1)=0, \, \, \,  \frac{\partial \tilde{\varphi}_1^{\dagger}}{\partial y} (1) = \frac{1}{Fe \bar{\phi}'(1)} \frac{\partial^2 \tilde{\varphi}_1^{\dagger}}{\partial y^2} (1), \, \, \, \frac{\partial \tilde{\varphi}_1^{\dagger}}{\partial y} (-1) = \frac{\tilde{\varphi}'_1 (-1)}{Fe \bar{\phi}'(-1)\tilde{\varphi}'_1 (-1)+\tilde{\varphi}''_1 (-1) } \frac{\partial^2 \tilde{\varphi}_1^{\dagger}}{\partial y^2} (-1).
	\end{equation}
\end{subequations}
More details regarding the derivation of boundary conditions here can be found in our previous work \citep{zhang2016weakly}.

\bibliography{BibRef.bib}
\end{document}